\documentclass[sigconf]{acmart}

\usepackage{enumitem} 
\usepackage{subcaption}
\usepackage{amsmath}
\usepackage{amsfonts}
\usepackage[ruled,vlined,linesnumbered]{algorithm2e}
\usepackage{xspace}
\usepackage{booktabs}
\usepackage{multirow}
\usepackage{graphicx}
\usepackage{booktabs}
\usepackage{pifont}
\usepackage{graphicx}
\usepackage{xcolor}

\newcommand{\sys}{AegisTS\xspace} 
\newcommand{\softbsubsec}[1]{\vspace{0.5em}\noindent\textbf{#1.}}

\newtheorem{definition}{Definition}

\AtBeginDocument{%
  }

\setcopyright{acmlicensed}
\copyrightyear{2018}
\acmYear{2018}
\acmConference[SIGMOD]{Make sure to enter the correct
  conference title from your rights confirmation email}{June 03--05,
  2018}{Woodstock, NY}
\begin{document}
\flushbottom


% \title{\sys: A Hierarchical Agent System with Reinforcement Learning for Multivariate Time Series Data Cleaning}
\title{\sys: A Hierarchical Agentic AI System with Reinforcement Learning for Multivariate Time Series Data Cleaning}


\author{Yuhan Shi}
% \orcid{0000-0001-5109-3700}
\affiliation{%
  \institution{Zhejiang University}
  \country{China}
}
\email{shiyuhan@zju.edu.cn}

% \hspace{0.1mm}

\author{Yuanyuan Yao}
\affiliation{%
  \institution{\mbox{National University of Singapore}}
  \country{Singapore}
}
\email{yy.yao@nus.edu.sg}

\author{Lu Chen}
\affiliation{%
  \institution{Zhejiang University}
  \country{China}
}
\email{luchen@zju.edu.cn}

\author{Mourad Khayati}
\affiliation{%
  \institution{University of Fribourg}
  \country{Switzerland}
}
\email{mourad.khayati@unifr.ch}

\author{Yushuai Li}
\affiliation{%
  \institution{Aalborg University}
  \country{Denmark}
}
\email{yusli@cs.aau.dk}

\author{Tianyi Li}
\affiliation{%
  \institution{Aalborg University}
  \country{Denmark}
}
\email{tianyi@cs.aau.dk}

\renewcommand{\shortauthors}{}

\begin{abstract}

Multivariate time series (MTS) are frequently affected by co-occurring quality issues, such as missing values, outliers, and constraint violations, which significantly undermine downstream analytics. Existing cleaning approaches fix only a limited set of such issues, making them ill-suited for scenarios where multiple quality problems arise simultaneously. Furthermore, these methods commonly depend on the availability of ground truth data or domain-specific rules, both of which are rarely accessible in real-world applications.

In this paper, we introduce \sys, an agent-driven reinforcement learning system designed to clean multiple data quality issues in MTS. We cast the cleaning process as a joint optimization problem that simultaneously handles quality issue order and cleaning model selection, allowing efficient navigation of the large space of possible cleaning pipelines. Our framework relies on a hierarchical agent architecture, where a high-level agent determines the order in which data quality issues should be processed, while a low-level agent identifies the most suitable cleaning method for each issue. To guide the agent toward an optimal cleaning pipeline, we propose a dual-stage reward mechanism that couples upstream (cleaning) and downstream performance, enabling effective optimization without relying on ground truth. Our experimental results show that \sys consistently outperforms existing methods, achieving up to 96\% improvement in data cleaning quality and 27\% improvement in downstream performance.

\end{abstract}

%\begin{CCSXML}
%<ccs2012>
%   <concept>
%       <concept_id>10002951.10002952.10003219.10003218</concept_id>
%       <concept_desc>Information systems~Data cleaning</concept_desc>
%       <concept_significance>500</concept_significance>
%       </concept>
%   <concept>
%       <concept_id>10002950.10003648.10003688.10003693</concept_id>
%       <concept_desc>Mathematics of computing~Time series analysis</concept_desc>
 %      <concept_significance>300</concept_significance>
%       </concept>
%\end{CCSXML}

%\ccsdesc[500]{Information systems~Data cleaning}
%\ccsdesc[300]{Mathematics of computing~Time series analysis}


\settopmatter{printacmref=false}
\renewcommand\footnotetextcopyrightpermission[1]{}
\pagestyle{plain}
\maketitle

\section{Introduction}

With the rapid proliferation of sensors, Internet of Things (IoT) devices, and large-scale information systems, massive amounts of time series data are continuously generated in real-world environments. While univariate time series describe the temporal dynamics of a single attribute, many real-world applications involve multivariate time series (MTS) that record multiple correlated attributes over time. By capturing both temporal patterns and cross-attribute dependencies, MTS provide rich information to support a wide range of tasks such as forecasting, classification, and clustering. 

% The value of MTS is fundamentally contingent on the quality of the underlying data, and in practice, this quality is far from guaranteed.
{The utility of MTS largely depends on the quality of the observations, which is often not guaranteed in practice.}
Real-world time series data are often plagued with various data quality issues, including missing values, duplicate values, outliers, and constraint violations. Missing values lead to incomplete observations, duplicate values produce redundant records at the same timestamp, outliers deviate significantly from normal patterns, and constraint violations occur when observations fail to satisfy expected temporal, cross-attribute, or speed constraints. 

Over the last decades, several specialized solutions for cleaning time series have been introduced \cite{DBLP:books/daglib/0005327,DBLP:journals/tkde/TakeuchiY06,kalman1960new,DBLP:journals/tsp/BreslerM86,DBLP:conf/icml/GetoorFKT01,dukhovny1990markov,DBLP:journals/pvldb/TuliCJ22,DBLP:journals/pvldb/ChenZMLDLHRLZ23,DBLP:conf/sigmod/SongZWY15,DBLP:journals/tods/SongGZWY21,DBLP:conf/dasfaa/YinYWHL18}. Among these, MissNet~\cite{DBLP:conf/kdd/ObataKMS24} and MPIN~\cite{DBLP:journals/pvldb/LiLLJPM23} address missing values through a state-space model and propagation; TranAD \cite{DBLP:journals/pvldb/TuliCJ22} and ImDiffusion \cite{DBLP:journals/pvldb/ChenZMLDLHRLZ23} detect and repair outliers based on Transformer reconstruction errors and diffusion-based generative modeling; and SCREEN~\cite{DBLP:conf/sigmod/SongZWY15} resolves constraint violations by keeping modifications minimal under speed constraints.
Despite their effectiveness, these techniques are generally designed to target a single category of data quality problem, making them inapplicable to real-world datasets where multiple quality issues tend to co-occur. 

A handful of algorithms tackle multiple quality issues simultaneously (see Table~\ref{tab:quality_issues}). The Kalman Filter (KF)~\cite{kalman1960new}, Hidden Markov Model (HMM)~\cite{gupta2012stock}, and EDITOR~\cite{li2026editor} all handle missing values and outliers. While the first two techniques rely on standard  filtering and probabilistic techniques, EDITOR distinguishes itself by using an advanced two-stage bidirectional approach with Temporal Convolutional Networks (TCN) and Graph Convolutional Networks (GCN) for context-aware repair. On the system side, Cleanits \cite{DBLP:journals/pvldb/DingWSLLG19} and Clean4TSDB \cite{DBLP:journals/pvldb/DingSWYWW24} (denoted as C4TSDB in Table~\ref{tab:quality_issues}) can address multiple data quality issues. Cleanits tackles missing values, outliers, and certain structural inconsistencies, whereas Clean4TSDB goes beyond basic imputation to detect and repair complex constraint violations, leveraging temporal and cross-attribute constraints for more comprehensive data correction.

These solutions fall short when applied to MTS with compound errors, for at least two reasons. First, they do not readily extend to quality issues beyond the specific combinations they were originally designed to address, restricting their generalizability to more diverse scenarios. Second, naively stacking multiple cleaning methods in sequence can easily break the inherent temporal and cross-attribute dependencies in the data. 
The lack of a unified cleaning solution for time series highlights the need for an adaptive, integrated system that can automatically choose and order the appropriate operators. Achieving this goal, however, is far from trivial and introduces two fundamental challenges.

\begin{figure}[t!]
 % \captionsetup[subfigure]{skip=2pt}
  \centering
  
  \begin{subfigure}{0.48\linewidth}
    \raggedright 
    \includegraphics[height=0.7\linewidth,keepaspectratio]{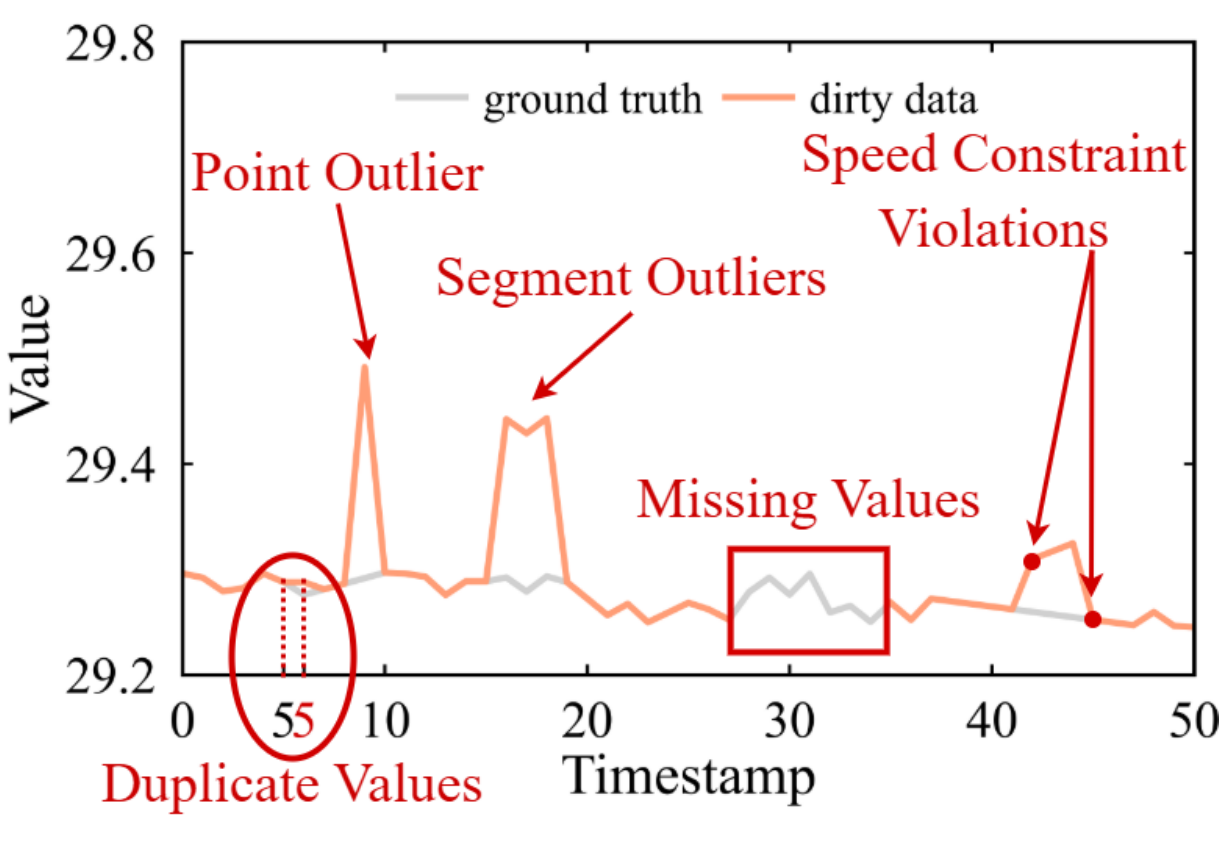}
    \caption{Quality issues in time series}
  \end{subfigure}% 
  \hspace{0.2cm} 
  \begin{subfigure}{0.48\linewidth}
    \raggedright 
    \includegraphics[height=0.7\linewidth,keepaspectratio]{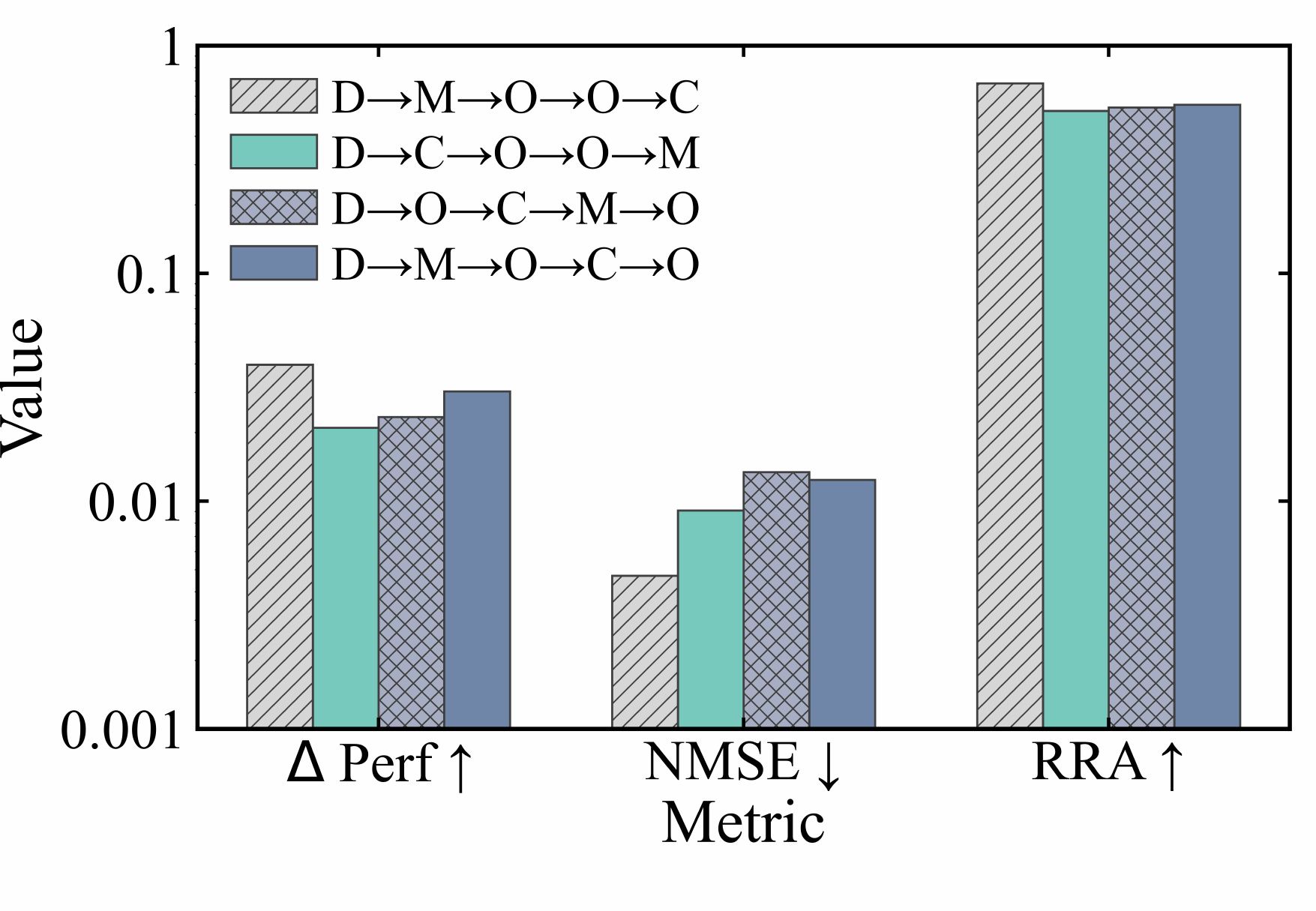}
    \caption{Execution order comparison}
  \end{subfigure}
  \vspace{-6mm}
%  \caption{Impact of cleaning execution order in the ETTh1 dataset.}
  \caption{Impact of cleaning execution order.}
  \Description{Case Study}
  \label{fig:case_study}
  \vspace{-4mm}
\end{figure}

First, designing an effective cleaning strategy requires more than simply selecting operators from a large candidate pool; it also requires determining the order in which they are applied. 
To illustrate this point, we evaluated different execution orders of cleaning methods on the widely used ETTh1 forecasting dataset. 
\autoref{fig:case_study}(a) shows a sample series that exhibits multiple co-occurring quality issues: duplicates (D), missing values (M), outliers (O), and constraint violations (C). Note that a cleaning sequence may invoke multiple operators of the same quality issue to handle distinct subtypes of errors. For instance, resolving both point outliers and segment outliers necessitates applying different outlier ($O$) operators. The results in~\autoref{fig:case_study}(b) show substantial performance variation across orderings, with $D \rightarrow M \rightarrow O \rightarrow O \rightarrow C$ yielding the best cleaning performance in terms of NMSE and RRA, as well as the highest downstream gain ($\Delta \text{Perf}$). {These results highlight the importance of execution order, but also the limitations of fixed-order strategies. The optimal strategy depends on the distribution and severity of quality issues, while each cleaning action alters the data-quality condition and affects subsequent decisions. Jointly optimizing quality issue selection, cleaning operators, and their execution order creates a large combinatorial search space, making effective strategy optimization challenging.}
% Given that time series can encompass over 400 distinct attributes~\cite{DBLP:conf/icde/KhayatiCTC25}, it is easy to conceive that the optimal ordering identified in one series may not transfer to another.
% In addition to the quality, the quest for the best execution order induces a combinatorial search space of complexity up to $\mathcal{O}(m^n)$, where $n$ denotes the number of quality issue types and $m$ the number of candidate methods per issue type, making the identification of the optimal cleaning strategy a non-trivial problem.

% Second, selecting appropriate cleaning operators typically requires ground truth. In particular, existing constraint-based cleaning methods, such as Clean4MTS\cite{DBLP:conf/icde/DingLWWS24} and MTSClean \cite{DBLP:journals/pvldb/DingSWWY24}, depend on inferring constraints from relatively clean data, implicitly assuming that the data is largely clean. However, in real-world applications, whether the time series are sufficiently clean is usually unknown, making it difficult to directly assess the effectiveness of different cleaning operators.
% Moreover, in the absence of ground truth, independently evaluating a single operator can be biased or even misleading, as its performance is often influenced by other unresolved data quality issues as well as the execution order of the cleaning operators. 

Second, the lack of clean ground-truth time series makes constraint acquisition and cleaning-effect evaluation challenging. 
Existing constraint-based methods~\cite{DBLP:conf/icde/DingLWWS24, DBLP:journals/pvldb/DingSWWY24} usually rely on constraints mined from relatively clean historical data or manually specified by domain experts. 
However, for dirty data with unknown constraints and no clean reference, obtaining reliable constraints becomes difficult. 
Furthermore, in the absence of ground-truth clean data, it is impossible to directly assess the quality of a cleaned series, forcing cleaning strategy optimization to depend on indirect quality indicators and downstream task feedback.

% To address those challenges, we devise an \tb{agent-driven} Hierarchical Reinforcement Learning (HRL) system that formulates multivariate time series cleaning as a sequential decision-making problem over an extensible space of cleaning operators. The system consists of three components: a preprocessing module for standardized data representation, a domain-specific operator repository for modeling cleaning operators and their dependencies, and \tb{HRL-based agents} with a dual-stage reward mechanism for automatically constructing effective dataset-specific cleaning pipelines. 

{

To address these challenges, we devise an agent-driven Hierarchical Reinforcement Learning (HRL) system that formulates multivariate time series cleaning as a sequential decision-making problem over an extensible space of cleaning operators. The system consists of two tightly coupled modules: a Data Quality Detector that standardizes the input data and characterizes its current data-quality condition, and a Cleaning Pipeline Generator that uses hierarchical agents and a dual-stage reward mechanism to iteratively select and compose suitable cleaning operators.
}To efficiently explore the combinatorial space induced by different cleaning operator choices and execution orders, the framework decomposes pipeline construction into two coordinated stages. A high-level policy first identifies the most critical data quality issue, and a low-level policy then selects and executes a corresponding cleaning operator. {After each operator is applied, the updated data is re-evaluated, and the resulting state is used to determine the next issue and operator.} By separating issue selection from operator instantiation, this design maps the original joint decision space over execution order and operator choice to a {step-wise} two-stage process, thereby reducing search computation cost from exponential to linear.
%, and enabling more efficient construction of dataset-specific cleaning pipelines. 

We further propose a dual-stage reward mechanism for learning without the clean ground truth. {Specifically, constraints are mined directly from dirty data, and their violation signals are used to derive state indicators and dense rewards.} Dense rewards provide step-wise feedback based on both reductions in data quality issue rates and performance gains of a lightweight model, while sparse rewards evaluate the overall cleaning strategy using a more complex model at the end of each iteration. By jointly exploiting upstream and downstream performance, this mechanism enables effective learning of high-quality cleaning policies without explicit supervision.

\begin{table}[t]
 \caption{Comparison of data quality issues addressed by different cleaning methods.}
 \vspace{-3mm}
  \label{tab:quality_issues}
  \centering
  \setlength{\tabcolsep}{3pt}
  \begin{tabular}{l|cccccc}
    \toprule
    Issue & KF & HMM & EDITOR & Cleanits & C4TSDB & \sys \\
    \midrule
    Duplicates & -- & -- & -- & -- & -- & \ding{51} \\
    Missing    & \ding{51} & \ding{51} & \ding{51} & \ding{51} & \ding{51} & \ding{51} \\
    Outliers   & \ding{51} & \ding{51} & \ding{51} & \ding{51} & -- & \ding{51} \\
    Constraint & -- & -- & -- & -- & \ding{51} & \ding{51} \\
    \bottomrule
  \end{tabular}
\end{table}

% The main contributions of this paper are summarized as follows:
% \begin{itemize}[topsep=0pt, partopsep=0pt]

%   \item We propose \sys ($\S$\ref{sec:framework}), the first {HRL system} for MTS cleaning, which formulates data cleaning as an automated principled composition of dataset-specific operator pipelines.

%    \item \sys builds a HRL model ($\S$\ref{sec:detector}), where a high-level agent prioritizes critical issues and guides a conditioned low-level agent to select appropriate repair operators. %, thereby mitigating error propagation.

%   \item We introduce a dual-stage reward mechanism that integrates upstream cleaning quality and downstream performance gains, enabling effective cleaning without requiring ground truth ($\S$\ref{sec:agent}).
  
% %  \item Extensive experiments on diverse real-world MTS datasets show that \sys significantly outperforms state-of-the-art baselines in downstream performance and data quality, while incurring a marginal one-time training computational overhead.
%     \item Extensive experiments ($\S$\ref{sec:eval}) on real-world MTS datasets show that \sys consistently outperforms state-of-the-art baselines in both data quality and downstream performance, while incurring only a minimal one-time training overhead.
% \end{itemize}

\par
\begingroup
\interlinepenalty=0
\clubpenalty=0
\widowpenalty=0
\displaywidowpenalty=0

\makeatletter
\@beginparpenalty=0
\@itempenalty=0
\@endparpenalty=0
\makeatother

The main contributions of this paper are summarized as follows:
\begin{itemize}[
  topsep=0pt,
  partopsep=0pt,
  itemsep=1pt,
  parsep=0pt
]
  \item We propose \sys ($\S$\ref{sec:framework}), the first HRL system for MTS cleaning, which formulates data cleaning as an automated process for composing dataset-specific operator pipelines in a principled manner.

  % \item \sys builds an HRL model ($\S$\ref{sec:detector}), where a high-level agent prioritizes critical issues and guides a conditioned low-level agent to select appropriate repair operators.

  % \item We introduce a dual-stage reward mechanism that integrates upstream cleaning quality and downstream performance gains, enabling effective cleaning without requiring ground truth ($\S$\ref{sec:agent}).

  \item We develop a Data Quality Detector ($\S$\ref{sec:detector}) that mines constraints from dirty data and quantifies issue severity to provide structured state signals.

  \item We design an HRL-based Cleaning Pipeline Generator ($\S$\ref{sec:agent}), where a high-level agent prioritizes quality issues and a conditioned low-level agent selects corresponding repair operators. A dual-stage reward mechanism jointly exploits cleaning-quality improvements and downstream performance, enabling effective learning without clean ground truth.

  \item Extensive experiments ($\S$\ref{sec:eval}) on real-world MTS datasets show that \sys consistently outperforms state-of-the-art baselines in both data quality and downstream performance, while incurring only a minimal one-time training overhead.
\end{itemize}
\endgroup

%The rest of this paper is organized as follows. Section~\ref{sec:back} introduces the preliminaries and problem definition. Section~\ref{sec:framework} overviews \sys, our HRL-based system for multivariate time series cleaning. Sections~\ref{sec:detector} and~\ref{sec:agent} present the technical details. Section~\ref{sec:eval} reports the experimental results, Section~\ref{sec:rel_work} reviews related work, and Section~\ref{sec:conclusion} concludes the paper.

\section{Preliminaries}
\label{sec:back}

In this section, we introduce the key concepts related to our holistic time series cleaning problem and formalize our problem definition.

\subsection{Data Quality Constraints}
\label{subsec:constraints}

\begin{definition}[Multivariate Time Series]
A multivariate time series is an ordered sequence of observations indexed by discrete timestamps. Throughout the paper, we use  $\mathbf{X} = \{\mathbf{x}_t\}_{t=1}^T \in \mathbb{R}^{T \times D}$ to denote an observed multivariate time series of length $T$, where each observation $\mathbf{x}_t \in \mathbb{R}^D$ is a $D$-dimensional vector representing the values of $D$ distinct attributes at time step $t$. 
\label{def:mts}
\end{definition}

To avoid ambiguity, we use the term attribute as the complete sequence of observations for a specific dimension, effectively forming a univariate time series. Accordingly, we use $\mathbf{x}^{(d)} \in \mathbb{R}^T$ to denote the complete time series of the $d$-th attribute and $x_{t,d}$ to denote its scalar observation at time $t$. Also, we distinguish the observed sequence $\mathbf{X}$, which is subject to potential data quality issues, from its corresponding unobserved ground-truth (clean) multivariate time series, denoted as $\mathbf{X}^* \in \mathbb{R}^{T \times D}$. To capture the dependencies that $\mathbf{X}^*$ adheres to, we identify two categories of data quality constraints:

\softbsubsec{Temporal Constraints (Column-wise)} A temporal constraint regulates the continuity of a single attribute $d$ across adjacent time steps, denoted as a quadruple $\sigma_{col} = (g, d, g_{min}, g_{max})$. Here, $g$ represents a temporal transition function designed to capture the allowable physical dynamics for a specific sensor (e.g., the step-wise rate of change $g = x_{t, d} - x_{t-1, d}$, intuitively referred to as \textit{Speed}). This constraint restricts the transition $g$ at any time step to the range $[g_{min}, g_{max}]$, formalized as:
\begin{equation}
    g_{min} \le g(x_{t, d}) \le g_{max}
\end{equation}
\noindent
where [$g_{min},g_{max}$] defines the permissible range for the temporal transition function $g$, capturing the allowable physical dynamics for a specific sensor. {The set of all such column-wise temporal constraints is denoted by $\Sigma_{col}$.}

\softbsubsec{Cross-attribute Constraints (Row-wise)} This constraint restricts the functional relationships among a specific subset of attributes at any given timestamp, which is represented as $\sigma_{row} = (f, \mathcal{V}, f_{min}, f_{max})$. Here, $\mathcal{V} \subseteq \{1, \dots, D\}$ denotes the indices of the attributes involved. 
The function $f: \mathbb{R}^{|\mathcal{V}|} \rightarrow \mathbb{R}$ is a multivariate polynomial function that maps the selected attribute values at a timestamp to a scalar consistency value, which is expected to fall within $[f_{min}, f_{max}]$ for valid observations.
%The function $f: \mathbb{R}^{|\mathcal{V}|} \rightarrow \mathbb{R}$ is a multivariate polynomial function that captures explicit algebraic interdependencies among these attributes. For any timestamp $t \in [1, T]$, the output of this polynomial mapping must satisfy:
\begin{equation}
    f_{min} \le f(\mathbf{x}_{t, \mathcal{V}}) \le f_{max}
\end{equation}
where $\mathbf{x}_{t, \mathcal{V}}$ is the vector of values for attributes indexed by $\mathcal{V}$ at time $t$, and $f_{min}$ and $f_{max}$ denote the lower and upper bounds of the permissible interval for the constraint function $f$, capturing explicit algebraic interdependencies among attributes. The set of all such row-wise cross-attribute constraints is denoted by $\Sigma_{row}$.

\begin{table}[t]
\centering
\small
\caption{Examples of MTS data quality constraints} \vspace{-3mm}
\label{tab:constraint_example}
\begin{tabular}{l}
\toprule
\ding{172} $\sigma_{\text{row}}:~
0 \leq Temp[t]-0.6Load[t]-0.3Vib[t]\leq 5$ \\
\ding{173} $\sigma_{\text{row}}:~
-3 \leq Press[t]-0.4Load[t]-0.2Vib[t]^2\leq 3$ \\
\ding{174} $\sigma_{\text{col}}:~
-0.5 \leq Temp[t+1]-Temp[t]\leq 0.5$ \\
\ding{175} $\sigma_{\text{col}}:~
-0.3 \leq Vib[t+2]-2Vib[t+1]+Vib[t]\leq 0.3$ \\
\bottomrule
\end{tabular}
\end{table}

\textbf{Example 1.}
Consider an equipment-monitoring MTS with oil temperature, motor
load, vibration, and pressure. As shown in
Table~\ref{tab:constraint_example}, row constraints
\ding{172}--\ding{173} describe cross-attribute dependencies at the
same timestamp, where the former is a first-order linear dependency
and the latter involves a higher-order polynomial term. Column
constraints \ding{174}--\ding{175} capture temporal speed and
acceleration patterns. Violations of these constraints indicate
abnormal observations that should be considered during cleaning.

The complete set of semantic rules encoding data quality is defined as the union of both constraint families, i.e.,  $\Sigma = \Sigma_{col} \cup \Sigma_{row}$. Rather than treating $\Sigma$ as prior domain knowledge, our approach seeks to automatically infer the concrete instantiations of these constraints, namely, the parameterization of $g$, $f$, and the bounding intervals, directly from the observed dirty data $\mathbf{X}$.

\subsection{Cleaning Pipelines}
\label{subsec:pipeline}

% In this work, we introduce an extensible set of $K$ specialized cleaning operators:
% \begin{equation}
%     \mathcal{O} = \{o_1, o_2, \dots, o_K\}
% \end{equation}
% where each operator $o_k \in \mathcal{O}$ constitutes a transformation mapping a corrupted data entry to its corrected counterpart. Throughout this paper, we fix $K = 48$, though the operator set can be extended.

% We use those operators to define a cleaning pipeline, which refers to an ordered sequence of operations designed to identify and remediate multiple data quality issues.

% \begin{definition}
% A cleaning pipeline $P$ of length $\mathcal{L}$ as an ordered sequence of operators drawn from $\mathcal{O}$. \textcolor{blue}{Since $o_{i_1}$ is the first operator executed on the data, standard mathematical notation for function composition dictates a right-to-left ordering:}
% \begin{equation}
%     % P = (o_{i_1}, o_{i_2}, \dots, o_{i_\mathcal{L}}), \quad \text{where } o_{i_l} \in \mathcal{O}
%     P = (o_{i_\mathcal{L}} \circ \dots \circ o_{i_2} \circ o_{i_1}), \quad \text{where } o_{i_l} \in \mathcal{O}
% \end{equation}

% \label{def:pipe}
% \end{definition}

We introduce an extensible set of $K$ specialized cleaning operators:
\begin{equation}
    \mathcal{B} = \{o_1, o_2, \dots, o_K\}
\end{equation}
where each operator $o_k \in \mathcal{B}$ is defined as a transformation function 
$o_k: \mathbb{R}^{T \times D} \rightarrow \mathbb{R}^{T \times D}$, 
which takes a multivariate time series as input and outputs its repaired version.
% where each operator $o_k \in \mathcal{B}$ is a transformation mapping a corrupted data entry to its corrected counterpart. 

%We fix $K = 48$ throughout this paper, though additional operators can be incorporated as needed. 
The operator library is extensible and can incorporate additional cleaning operators as needed. These operators serve as the building blocks of a cleaning pipeline aimed at detecting and resolving multiple data quality issues.

\begin{definition}
A \emph{cleaning pipeline} $P$ of length $\mathcal{L}$ is a sequence of operators drawn from $\mathcal{B}$. Formally, 
\begin{equation}
    P = (o_{i_1}, o_{i_2}, \dots, o_{i_\mathcal{L}}), \quad \text{where } o_{i_l} \in \mathcal{B}
\end{equation}
so that $o_{i_1}$ is the first operator applied to the input data.
\label{def:pipe}
\end{definition}

% The execution of a cleaning pipeline $P$ on dirty data $\mathbf{X}$ is equivalent to the sequential composition of its constituent operators. To formally describe the sequential execution of pipeline $P$, let $\mathbf{X}^{(0)} = \mathbf{X}$ denote the initial dirty data \mkh{we cannot use two different symbols to refer to the same entity}.  The intermediate data state after applying the $l$-th operator is $\mathbf{X}^{(l)} = o_{i_l}(\mathbf{X}^{(l-1)})$. The final cleaned data is obtained through the composition of all $L$ operators. Thus, the complete pipeline execution can be expressed as: \mkh{you introduce $\mathbf{X}^{(0)}$ and $\mathbf{X}^{(l)} $ and do not use any of them} \textcolor{blue}{(I use $\mathbf{X}^{(0)}$ and $\mathbf{X}^{(l)} $ in the algorithm). Then, introduce them when needed!}

The execution of a cleaning pipeline $P$ on dirty data $\mathbf{X}$ is equivalent to the composition of its constituent operators. Formally, the final cleaned data is obtained by applying all $L$ operators in order, which can be expressed as a function $\mathcal{P}(\mathbf{X})$:
\begin{equation}
    \mathbf{X}_{\text{cleaned}} = \mathcal{P}(\mathbf{X}) = (o_{i_\mathcal{L}} \circ \dots \circ o_{i_2} \circ o_{i_1})(\mathbf{X})
    \label{eq:pipeline_composition}
\end{equation}

% The execution order of operators is critical to cleaning performance. As shown in \autoref{fig:figure2} \mkh{should we move this experiment to the intro?}, we compare two different sequences of the same four operators (\textit{Interpolation}, \textit{AR}, \textit{SVR}, and \textit{SCREEN}) using the ETTh1 dataset. Simply swapping the positions of the two outlier correctors ($o_{AR}$ and $o_{SVR}$) causes the MSE to rise from $0.0047$ to $0.0124$. More importantly, the downstream model performance gain ($\Delta$ Perf) drops from 0.0402 to 0.0247, representing a substantial 38.6\% reduction in cleaning utility. This evidence suggests that cleaning operators are generally non-commutative. Because the effectiveness of an operator depends on the data state $\mathbf{X}^{(l-1)}$ produced by its predecessors, selecting the right operators is not enough. A robust system must also find the optimal execution sequence to ensure the best cleaning results.

\begin{figure*}[tb]
  \centering
  \includegraphics[width=0.9\textwidth,height=\linewidth,keepaspectratio]{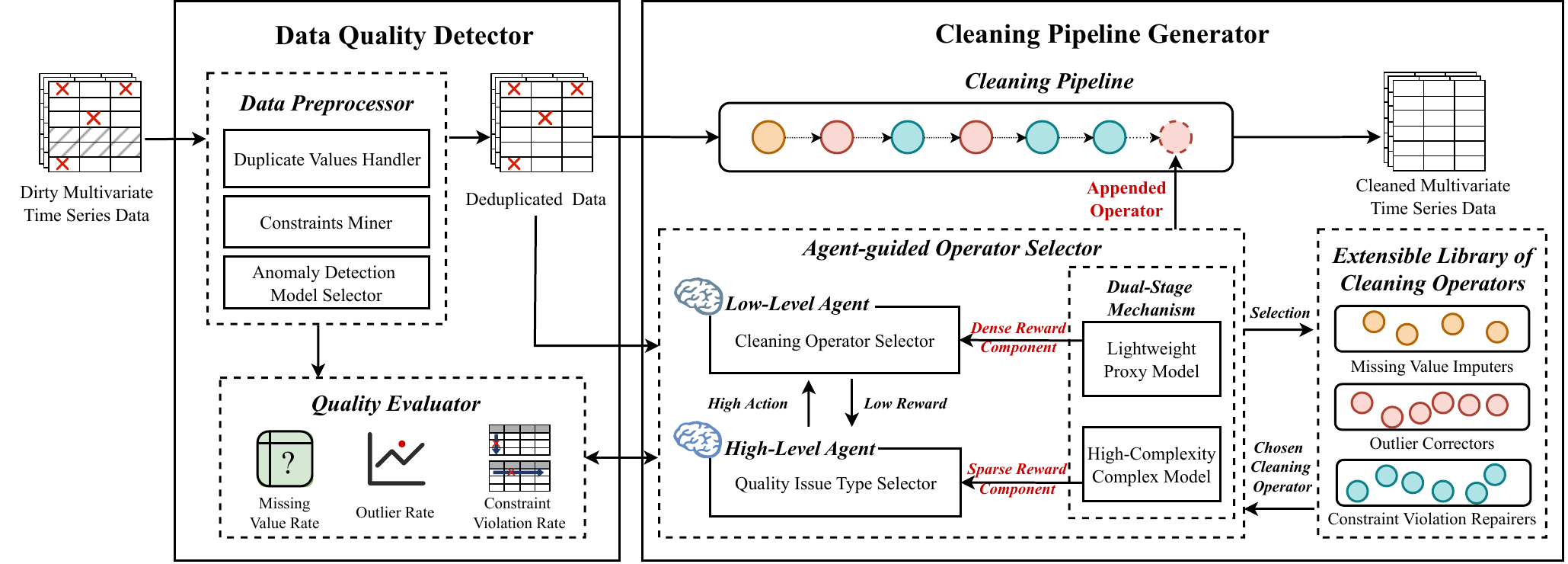} %\vspace{-7mm}
  \caption{Overall framework of \sys.} 
  \Description{Overall framework of \sys, including Data Quality Detector and Cleaning Pipeline Generator with HRL and dual-stage rewards.}
  \vspace{-4mm}
  \label{fig:sys}
\end{figure*}

\subsection{Problem Statement}
\label{subsec:problem-def}

Unlike existing approaches that rely on ground-truth data, we cast holistic data cleaning as a task-driven combinatorial optimization problem. Our framework navigates a large space of candidate operator sequences to discover optimal cleaning pipelines that produce cleaned datasets maximizing downstream model performance, while reducing violations of the intrinsic structural constraints $\Sigma$ inferred directly from the dirty data.
%while strictly preserving intrinsic structural constraints inferred directly from the dirty data, thereby eliminating any reliance on reference data.

\noindent\textbf{Problem Definition.} Let $\mathbf{X}$ be a dirty multivariate time series dataset, $\mathcal{B}$ be a set of candidate cleaning operators, and $\mathcal{M}$ be a target downstream task model. Given the inferred constraint family $\Sigma$ defined above, our goal is to discover an optimal cleaning pipeline $P^* = (o_{i_1}, o_{i_2}, \dots, o_{i_\mathcal{L}})$, where $o_{i_j} \in \mathcal{B}$, that maximizes the downstream task performance while reducing the violation degree of $\Sigma$. Formally, we aim to solve:
% while strictly satisfying a set of robustly mined intrinsic structural constraints $\Sigma$. Formally, we aim to solve:
\begin{align}
    P^* = \underset{P \in \Omega}{\arg\max} \;\; \Phi\big(\mathcal{M},\, \mathcal{P}(\mathbf{X})\big)
\end{align}
where $\Phi(\cdot,\cdot)$ is the evaluation metric of the downstream model $\mathcal{M}$, 
$\Omega$ is the space of all admissible repair operator sequences, 
and $\mathcal{P}(\mathbf{X}) = \mathbf{X}_{\text{cleaned}}$ is the cleaned dataset obtained by applying pipeline $P$ to the dirty data $\mathbf{X}$. Note that since the pipeline utilizes third-party black-box operators, we do not impose a strict binary constraint (i.e., $\mathcal{P}(\mathbf{X}) \models \Sigma$). Instead, the operators implicitly work towards satisfying the integrity constraint set $\Sigma$ on a best-effort basis, guided by the downstream metric.

% Unlike traditional methods that aim to approximate an unknown ground truth, we formulate this automated cleaning process as a \textit{task-driven combinatorial optimization problem}. Specifically, the generated pipeline $P$ must maximize the performance of the downstream model $\mathcal{M}$ on the cleaned data, while strictly preserving a set of intrinsic structural constraints $\Sigma$. Crucially, to avoid the reliance on pristine reference data, this constraint set $\Sigma$ is robustly mined directly from the dirty dataset $\mathbf{X}$ beforehand. 

% Ultimately, finding $P$ requires navigating a massive combinatorial action space to discover synergistic operator sequences, which is designed to elevate data quality while ensuring the downstream task achieves maximum performance. The final outputs include the derived cleaning pipeline $P$, and the resulting cleaned dataset $\mathbf{X}_{clean}$.

\sloppy

\section{\sys Overview}
\label{sec:framework}

% \begin{figure*}[htb!]
%   \centering
%   \includegraphics[width=\textwidth,height=\linewidth,keepaspectratio]{figures/framework.pdf} \vspace{-3mm}
%   \caption{The overall framework of \sys.} 
%   \Description{Overall framework of \sys, including Data Quality Detector and Cleaning Pipeline Generator with hierarchical RL and dual-stage rewards.}
% %  \vspace{-4mm}
%   \label{fig:sys}
% \end{figure*}

% \begin{table*}[t]
%   \caption{Comparison of data quality issues addressed by different cleaning methods} 
%   %\vspace{-4mm}
%   \label{tab:quality isuues}
%   \centering
%   \begin{tabular}{lccccccccccl}
%     \toprule
%     Quality Issue & Kalman Filter \cite{kalman1960new} & HMM \cite{gupta2012stock} & EDITOR \cite{li2026editor} & Cleanits \cite{DBLP:journals/pvldb/DingWSLLG19} & Clean4TSDB \cite{DBLP:journals/pvldb/DingSWYWW24} & \sys \\

%     \midrule
%     Duplicate Values & -- & -- & -- & -- & -- & \ding{51}  \\
%     Missing Values & \ding{51} & \ding{51} & \ding{51} & \ding{51} & \ding{51} & \ding{51}  \\
%     Outliers & \ding{51} & \ding{51} & \ding{51} & \ding{51} & -- & \ding{51}  \\
%     Constraint Violations & -- & -- & -- & -- & \ding{51} & \ding{51}  \\
%     \bottomrule
%   \end{tabular}
% \end{table*}

The goal of \sys is to take a dirty multivariate time series and automatically generate an optimal cleaning pipeline that maximizes downstream task performance while preserving intrinsic structural constraints. Two salient design decisions motivate the architecture: (a) cleaning is formulated as a sequential decision-making problem guided by reinforcement learning, as opposed to a single-shot selection process, and (b) the search operates over full operator sequences rather than individual operator choices in isolation. \autoref{fig:sys} presents the framework's main components and their interactions.
The framework operates through two tightly coupled modules: the Data Quality Detector and the Cleaning Pipeline Generator. We discuss each in turn.

\subsection{Data Quality Detector}
The Data Quality Detector serves as the observation function of the RL environment. Given a dirty multivariate time series, it characterizes the data quality issues present and encodes them into a structured state representation consumable by the RL agent. The process begins with the Data Preprocessor, which removes duplicate records and simultaneously initializes two parallel components: a Constraint Miner, which extracts intrinsic structural rules directly from the dirty data, and an Anomaly Detection Model Selector, which constructs appropriate outlier detection models for the dataset. The deduplicated data are maintained as the current data instance in the RL environment, rather than being directly flattened as the policy input. The Quality Evaluator summarizes the quality condition of this data instance using three metrics: the Missing Value Rate, the Outlier Rate derived from the selected anomaly detection models, and the Constraint Violation Rate assessed against the mined constraints. These quality metrics constitute the agent-observable quality state, while the current data instance is retained by the environment for executing subsequent cleaning actions. The mined constraints are further propagated to downstream operators to guide rule-aware cleaning decisions.

% Building on this, the Quality Evaluator quantifies data quality along three dimensions: the Missing Value Rate, the Outlier Rate derived from the selected anomaly detection models, and the Constraint Violation Rate assessed against the mined constraints. The deduplicated data and these quality metrics are consolidated into a unified state representation that provides the RL agent with an initial characterization of the current data quality, while the fetched constraints are propagated to downstream operators to guide subsequent cleaning decisions.

\vspace{-2mm}
\subsection{Cleaning Pipeline Generator}
The Cleaning Pipeline Generator constitutes the core decision-making module of the framework, orchestrating the step-wise selection and execution of cleaning operators. %orchestrating the selection and sequencing of cleaning operators. 
To support diverse cleaning needs, it maintains an extensible set of operators spanning three functional categories: Missing Value Imputers, Outlier Correctors, and Constraint Violation Repairers. Operator selection is governed by an HRL architecture: At each step, the High-Level Agent first selects a quality issue category under the current data state, and the Low-Level Agent then selects a specific operator from the corresponding category. After execution, the updated data is re-evaluated to guide the next decision.
% a High-Level Agent determines which category of quality issue to address at each step, while a Low-Level Agent selects the most appropriate operator within the identified category.

Pipeline construction is guided by a Dual-Stage Reward Mechanism. A Lightweight Model provides dense, immediate rewards after each individual cleaning operation, accelerating policy exploration and convergence toward high-quality sequences. A High-Complexity Model is invoked once the complete pipeline is assembled, delivering sparse rewards reflective of true downstream task performance. By integrating dense and sparse feedback, this mechanism enables the learned policy to effectively balance data restoration fidelity with downstream utility. The framework operates as a closed-loop system, iteratively observing the current data quality state, selecting strategic cleaning actions, and refining its policy through downstream task feedback, ultimately synthesizing a customized pipeline that transforms corrupted multivariate time series into high-quality data ready for downstream analytics.

\section{Data Quality Detector}
\label{sec:detector}

The first component of \sys includes two central modules acting sequentially.  The Data Preprocessor standardizes the raw input, while the Quality Evaluator quantifies issue severity to guide downstream repair prioritization.

\subsection{Quality Evaluator}
\label{sec:quality_rates}

Although we identify four types of quality issues, we compute three quality issue rates: the missing rate $\mathbf{r}_{missing}$, the outlier rate $\mathbf{r}_{outlier}$, and the violation rate $\mathbf{r}_{violation}$. We use only three rates, as duplicate values are converted into missing values. Specifically, each rate is calculated as the ratio of data points affected by a specific issue to the total dataset size.
Through the above quantified metrics, we can evaluate the relative severity of different issues. By identifying the most prominent data quality issue, we can prioritize its resolution. Consequently, these rates act as critical indicators that guide the optimal order of cleaning operations.

\subsection{Data Preprocessor}

%To establish a foundation for quantifying data quality issues, we sequentially handle duplicates, mine robust constraints, and select anomaly models for outlier detection.

To establish a foundation for the cleaning operators, we present the different techniques we apply to handle the quality issues.

\subsubsection{Constraint Miner}
Conventional constraint mining for time series assumes relatively clean data \cite{DBLP:conf/icde/DingLWWLW24,DBLP:journals/pvldb/DingSWWY24}. In practice, anomalies can severely distort regression-based constraint induction. To ensure robustness, our mining process operates directly on dirty data $\mathbf{X}$ and combines MAD-based statistics for temporal constraints with feature-residual pruning for cross-attribute dependencies. 

\softbsubsec{Temporal Constraint Mining} For column-wise temporal constraints $\Sigma_{col}$, we mine three types:  Speed (first-order difference), Acceleration (second-order difference), and Variance (local variance within a sliding window). For each attribute $d$, applying function $g$ (cf. Section~\ref{sec:back}) to all valid samples yields an empirical value set $\mathbf{G}$ (e.g., all speed values for $d$).
Rather than relying on mean and standard deviation, both sensitive to extreme outliers, we adopt the Median Absolute Deviation (MAD) to derive a robust bounding interval: $[g_{min}, g_{max}] =$
\begin{equation}
  \left[ \mathrm{median}(\mathbf{G}) - k \cdot \sigma_{robust}, \mathrm{median}(\mathbf{G}) + k \cdot \sigma_{robust} \right]
\end{equation}
\noindent
where $\sigma_{robust} = 1.4826 \cdot \mathrm{median}(|\mathbf{G} - \mathrm{median}(\mathbf{G})|)$ is the MAD-based robust estimate of the standard deviation. The factor 1.4826 is the standard normal-consistency correction that rescales MAD to be comparable to the standard deviation under the normality assumption. We set k=3 to obtain a robust analogue of the conventional three-sigma interval.

% where $\sigma_{robust} = 1.48 \cdot \mathrm{median}(|\mathbf{G} - \mathrm{median}(\mathbf{G})|)$ and $k$ is set according to the three-sigma rule. This estimation is applied independently across all attributes and constraint types, tightly capturing normal physical inertia while suppressing erratic corruptions.

\softbsubsec{Cross-attribute Constraint Mining} To derive row-wise cross-attribute constraints $\Sigma_{row}$, we introduce a robust, lightweight procedure that integrates correlation-based feature screening with polynomial modeling.
We first filter a relevant subset via correlation and redundancy analysis. Multivariate polynomial relationships are then modeled on this refined subset. Candidate constraints are retained only if they exhibit strong goodness-of-fit and low residual variance. Final tolerance bounds are derived from empirical residual quantiles, followed by pruning of insignificant terms, yielding compact, stable, and interpretable constraint forms.

\subsubsection{Anomaly Detection Model Selection}  
%We adopt FMMS \cite{DBLP:conf/smc/ZhangWXZ22} to automatically select the top-$k$ anomaly detection models best suited to the target time series. By mapping meta-features (e.g., trend, seasonality) to the most suitable detection algorithms, FMMS ensures precise model–data alignment, substantially improving detection accuracy over a single default model. The selected models produce anomaly scores independently, which are then averaged into a unified sequence; timestamps exceeding a predefined threshold are flagged as anomalies.

We adopt FMMS \cite{DBLP:conf/smc/ZhangWXZ22} following the benchmark study \cite{DBLP:journals/tmlr/ZhouBL26}, which identifies FMMS as an effective model selection method for time series anomaly detection. FMMS selects the top-$k$ models for the target time series by mapping meta-features (e.g., trend and seasonality) to suitable detection algorithms. The selected models produce anomaly scores independently, which are then averaged into a unified sequence; timestamps exceeding a predefined threshold are flagged as anomalies.

\subsubsection{Duplicate Values Handler} 
We remove duplicate timestamps by retaining only the first occurrence, producing a strictly increasing time series. The sampling interval is estimated as the mode of adjacent time differences, enabling alignment to a regular time grid. Missing grid points are marked as missing values, yielding a regular multivariate time series for subsequent quality detection.

\section{Agent-guided Operator Selector}
\label{sec:agent}

%Traditional cleaning pipelines \cite{feurer2015efficient} rely on static selection \mkh{of what?}, leading to two critical flaws: (1) open-loop execution that ignores distribution shifts caused by prior operations, causing error accumulation without feedback; and (2) combinatorial explosion in operator enumeration, making exhaustive search infeasible. In contrast, 
%As mentioned earlier, we reformulate cleaning as a closed-loop RL problem with a hierarchical architecture. %Our \textcolor{blue}{Markov Decision Process (MDP)} -based agents
Our agent-based solution dynamically selects operators conditioned on the current data state, enabling adaptive correction of earlier errors. To avoid combinatorial search, we introduce a two-layer hierarchy: a high-level agent diagnoses quality issues, while a low-level agent selects specific operators. This decomposition converts the joint optimization into tractable sub-tasks. A dual-stage reward further balances issue resolution and downstream task performance.

\subsection{Hierarchical Agent Architecture}

To address the combinatorial explosion in cleaning pipeline search space, we propose a two-layer HRL architecture that decouples issue identification from repair execution. This decomposition transforms an intractable joint search into structured sub-tasks, enabling efficient learning in large state–action spaces.

\softbsubsec{High-Level Agent (Strategic Issue Prioritization)} The high-level agent selects a discrete action $\mathcal{A}^H = \{\text{M}, \text{O}, \text{C}, \text{F}\}$, where \text{M}, \text{O}, \text{C}, and \text{F} denote missing-value repair, outlier correction, constraint-violation resolution, and termination of the cleaning pipeline generation, respectively. This choice determines which quality issue to address next.
Its state representation captures four critical dimensions. First, Urgency ($i_{dom}$) is a one-hot indicator of the most severe issue (i.e., the maximum among the three quality metrics), enabling priority-driven correction. Second, Memory ($a_{prev}^H$) is a one-hot encoding of the previous action, helping prevent redundant loops and promoting coherent action sequences. Third, Cost–Benefit Awareness ($p_{lite}$) reflects downstream task performance from a lightweight model, signaling when further cleaning yields diminishing returns. Finally, Budget Awareness ($l$) is a normalized step count (current step divided by the maximum budget), guiding timely termination.
The resulting state vector is $s^H = [i_{dom},\ a_{prev}^H,\ p_{lite},\ l]$. This design enables the high-level agent to dynamically prioritize, adapt, and terminate, effectively closing the loop between cleaning actions and downstream utility.

% The high-level agent selects a discrete action $ \mathcal{A}^H = \{\text{M}, \text{O}, \text{C}, \text{F}\}$ (\text{M}, \text{O}, \text{C}, and \text{F} denote missing-value repair, outlier correction, constraint-violation resolution, and termination of the cleaning process, respectively), determining which quality issue to address next. Its state representation captures four critical dimensions. First, \textit{Urgency} ($i_{dom}$): is a one-hot indicator of the most severe issue (max of three quality metrics), enabling priority-driven correction. Second, 
% \textit{Memory} ($a_{prev}^H$): is a one-hot encoding of the prior action, preventing redundant loops and promoting coherent sequences. Third, \textit{Cost-benefit awareness} ($p_{lite}$): is the downstream task performance from a lightweight model, signaling when further cleaning yields diminishing returns. And Last, \textit{Budget awareness} ($l$): is a normalized step count (current step / max budget), guiding timely termination. The resulting state vector is: $s^H = [i_{dom},\ a_{prev}^H,\ p_{lite},\ l]$. This design enables the high-level agent to dynamically prioritize, adapt, and terminate, closing the loop between cleaning actions and downstream utility.

\softbsubsec{Low-Level Agent (Operator Selection)} Given a high-level action $a^H \in \{\text{M}, \text{O}, \text{C}\}$ (issue type), the low-level agent selects a concrete cleaning operator $a^L \in \mathcal{A}^L$, where $\mathcal{A}^L$ corresponds to the sub-library of candidate tools (e.g., imputers for missing values). To prevent policy interference across distinct quality issues, we maintain independent Q-tables per high-level action, enabling specialized learning of category-specific repair logic.

The low-level state representation captures three complementary perspectives:
i) Goal Alignment ($g_{ind}$): a one-hot mask derived from $a^H$ that filters irrelevant operator sub-libraries, ensuring focus on the targeted issue;
ii) Data Distribution ($d_{stat}$): statistical features (e.g., skewness, sparsity, variance) that characterize the data profile and support informed operator matching; and
iii) Temporal Dynamics ($m_{struct}$): time-series-specific metrics (e.g., stationarity scores) that guide operator selection to preserve signal integrity and minimize distortion.
The resulting state vector is $s^L = [g_{ind},\ d_{stat},\ m_{struct}]$. This design enables the low-level agent to jointly consider what to fix (via $g_{ind}$) and how to fix it (via $d_{stat}$ and $m_{struct}$), enabling precise, context-aware operator selection.

\subsection{Dual-Stage Reward Mechanism}
\label{subsec:reward}

To enable dense feedback without prohibitive computational costs, we decouple reward generation: a lightweight proxy model provides dense intermediate rewards, while the target complex model supplies sparse terminal rewards. This dual-stage design aligns step-wise cleaning trajectories with end-to-end objectives.

\subsubsection{Low-Level Dense Reward}

The low-level reward evaluates four complementary dimensions. First, Temporal Smoothness ($\mathcal{R}_{structure}^{dense}$) penalizes spikes and jitter to preserve signal coherence. Second, the Modification Constraint ($\mathcal{R}_{distance}^{dense}$) measures the RMS difference between pre- and post-cleaning data, discouraging excessive tampering. Third, Local Effectiveness ($\mathcal{R}_{local}^{dense}$) captures the direct reduction in the targeted issue rate (e.g., a decrease in missing values). Finally, to align local cleaning with the ultimate goal without the prohibitive computational bottleneck of complex downstream model evaluation, Task Alignment ($\mathcal{R}_{lite}^{dense}$) reflects the performance gain of a lightweight model trained for the same downstream task, enabling rapid task-oriented feedback. The aggregated dense reward is defined as follows:

% Finally, Task Alignment ($\mathcal{R}_{lite}^{dense}$) reflects the performance gain of the downstream model, enabling rapid task-oriented feedback. 

\begin{equation}
    \mathcal{R}_{low}^{dense} = \mu_1 \mathcal{R}_{structure}^{dense} - \mu_2 \mathcal{R}_{distance}^{dense} + \mu_3 \mathcal{R}_{local}^{dense} + \mu_4 \mathcal{R}_{lite}^{dense}
    \label{eq:low_dense}
\end{equation}

Each term is normalized to stabilize the optimization. This joint design prioritizes data repairs that maximize downstream utility while preserving data integrity.

% \begin{itemize}
%     \item \textbf{Temporal smoothness $\mathcal{R}_{structure}^{dense}$:} It penalizes spikes and jitters to preserve signal coherence.
%     \item \textbf{Modification constraint $\mathcal{R}_{distance}^{dense}$:} It is RMS difference between pre- and post-cleaning data, discouraging excessive tampering.
%     \item \textbf{Local effectiveness $\mathcal{R}_{local}^{dense}$:}  It is a direct reduction in the targeted issue rate (e.g., missing value decrease).
%     \item \textbf{Task alignment $\mathcal{R}_{lite}^{dense}$:}  It is a performance gain of a lightweight downstream model, enabling rapid task-oriented feedback.
% \end{itemize}

\subsubsection{High-Level Reward}

The high-level agent receives a composite reward consisting of dense intermediate feedback and sparse terminal evaluation.

\paragraph{Dense Reward:}

At each non-terminal step, the high-level dense reward integrates four complementary signals. The Execution Feedback Propagation ($\mathcal{R}_{low}^{dense}$ in Equation~\ref{eq:low_dense}) propagates fine-grained low-level outcomes, favoring issue types that yield effective local repairs. The Global Quality Improvement ($\mathcal{R}_{quality}^{dense}$) reflects the reduction in the overall data quality issue rate, guiding macroscopic cleaning progress. The Computational Cost Penalty ($\mathcal{R}_{cost}^{dense}$) accounts for lightweight model execution time, balancing quality gains against efficiency. Finally, the Stagnation Penalty ($\mathcal{R}_{penalty}^{dense}$) imposes a discrete penalty for redundant operations (e.g., applying the same operator without improvement), breaking potential infinite loops and preventing conservative deadlock.
The aggregated dense reward is defined as follows:
\begin{equation}
    \mathcal{R}_{high}^{dense} = \lambda_1 \mathcal{R}_{low}^{dense} + \lambda_2 \mathcal{R}_{quality}^{dense} - \lambda_3 \mathcal{R}_{cost}^{dense} - \mathcal{R}_{penalty}^{dense}
    \label{eq:high_dense}
\end{equation}
 
The first three components are normalized for stable optimization; $\mathcal{R}_{penalty}^{dense}$ remains unnormalized to strictly outweigh incremental gains, avoiding any deadlocks.

\paragraph{Sparse Reward:}
At episode termination (action \text{F} or budget exhaustion), the agent receives:
\begin{equation}
    \mathcal{R}_{high}^{sparse} =  M_{\text{complex}}(X_{\text{cleaned}}) - M_{\text{complex}}(X)
\end{equation}
where $M_{\text{complex}}(\cdot)$ is the target downstream model performance.

This delayed terminal reward propagates backward via Q-learning, anchoring the policy to true end-to-end objectives.
Thus, the dual-stage design achieves efficiency via dense lightweight-model feedback and fidelity via sparse complex-model supervision.

\begin{algorithm}[tb]
\small \DontPrintSemicolon \SetAlgoLined
\caption{Iterative Training of Pipeline Generation}
\label{alg:hierarchical_cleaning}

\KwIn{Dirty data $\mathbf{X}$; operator library $\mathcal{B}$; quality evaluator $E$; lightweight model $M_{lite}$; complex model $M_{complex}$; episodes $N$; max pipeline length $\mathcal{L}_{\max}$}
\KwOut{Trained high-level agent $Agent_H(\theta_H)$ and low-level agent $Agent_L(\theta_L)$}

Initialize $M_{lite}$, $M_{complex}$, and parameters $\theta_H, \theta_L$\;

\For{$n \leftarrow 1$ \KwTo $N$}{
  $(\mathbf{X}^{(0)}, P, l) \leftarrow (\mathbf{X}, \emptyset, 0)$\;
  
  \While{$l < \mathcal{L}_{\max}$}{
    $\mathbf{Rates}^{(l)} \leftarrow E(\mathbf{X}^{(l)})$\;
    
    $a^H \leftarrow Agent_H(\textnormal{ExtractHighState}(\mathbf{X}^{(l)}, \mathbf{Rates}^{(l)}))$\;
    \lIf{$a^H = \textnormal{F}$}{\textbf{break}}
    
    $a^L \leftarrow Agent_L(\textnormal{ExtractLowState}(\mathbf{X}^{(l)}, a^H))$\;
    
    $(\mathbf{X}^{(l+1)}, P) \leftarrow (o(\mathbf{X}^{(l)}), P \oplus o) \textnormal{ where } o = \mathcal{B}(a^H, a^L)$\;
    
    $\mathbf{Rates}^{(l+1)} \leftarrow E(\mathbf{X}^{(l+1)})$\;
    $\mathcal{R}_{lite}^{dense} \leftarrow M_{lite}(\mathbf{X}^{(l+1)})$ \;
    
    Update $\theta_L$ using $\mathcal{R}_{low}^{dense}(\mathbf{Rates}^{(l+1)}, \mathcal{R}_{lite}^{dense})$ \;
    Update $\theta_H$ using $\mathcal{R}_{high}^{dense}(\mathbf{Rates}^{(l+1)}, \mathcal{R}_{lite}^{dense})$\;
    
    $l \leftarrow l + 1$\;
  }
  
  Update $\theta_H$ with $\mathcal{R}_{high}^{sparse} \equiv M_{complex}(\mathbf{X}^{(l)}) - M_{complex}(\mathbf{X})$\;
}

\Return{$Agent_H$, $Agent_L$}\; 
\end{algorithm} 

\subsection{Training and Optimization}

\subsubsection{Training Procedure.}
\label{sec:training_process}

Algorithm~\ref{alg:hierarchical_cleaning} 
%\mkh{we don't explain how this algorithm is used for cleaning! Can you add another cleaning algorithm that calls Alg1?} 
summarizes the reinforcement learning procedure for jointly training the high-level and low-level policies to generate a data cleaning pipeline. It takes as input the dirty dataset $\mathbf{X}$, the candidate operator pool $\mathcal{B}$, the quality evaluator $E$, the lightweight model $M_{lite}$, the downstream complex model $M_{complex}$, and the training hyperparameters $N$ and $\mathcal{L}_{\max}$. 

% The procedure starts by initializing $M_{lite}$, $M_{complex}$, and the policy parameters $\theta_H$ and $\theta_L$ for the two agents (Line 1).
% The training proceeds over $N$ episodes (Lines 2--16), where each episode resets the environment and pipeline $P$, restoring $\mathbf{X}^{(0)} \leftarrow \mathbf{X}$ and resetting the step counter $l$ (Line 3). As the agents sequentially apply operators, we use $\mathbf{X}^{(l)} = o_{i_l}(\mathbf{X}^{(l-1)})$ to denote the intermediate data state after applying the $l$-th operator. At each step $l$, the evaluator $E$ computes quality issue rates (Line 5). The high-level agent selects a high-action $a^H$ based on the extracted high-level state (Line 6); if $a^H = \textnormal{F}$ (\textnormal{F} denotes the termination of the cleaning pipeline generation in this episode \textcolor{blue}{where $\textnormal{F}$ is a termination action autonomously selected by the high-level agent when the marginal performance gain no longer justifies the additional computational cost.}), the pipeline building terminates (Line 7). Otherwise, the low-level agent selects a specific operator $a^L$ based on the extracted low-level state (Line 8), and the cleaning operator $o$ is executed, generating $\mathbf{X}^{(l+1)}$ and appending $o$ to $P$ (Line 9).

The procedure starts by initializing $M_{\text{lite}}$, $M_{\text{complex}}$, and the policy parameters $\theta_H$ and $\theta_L$ for the two agents (Line 1). Training then runs for $N$ episodes (Lines 2–16). At the start of each episode, the environment and pipeline $P$ are reset, setting $\mathbf{X}^{(0)} \leftarrow \mathbf{X}$ and initializing the step counter $l$ (Line 3). As operators are applied sequentially, the intermediate data state is denoted by $\mathbf{X}^{(l)} = o_{i_l}(\mathbf{X}^{(l-1)})$, representing the result after the $l$-th operator.
At each step $l$, the evaluator $E$ computes the quality issue rates (Line 5). The high-level agent then selects an action $a^H$ based on the high-level state (Line 6). If $a^H = \textnormal{F}$, a termination action chosen when the marginal performance gain no longer justifies the additional computational cost, the pipeline construction stops (Line 7). Otherwise, the low-level agent selects a specific operator $a^L$ based on the low-level state (Line 8), after which the corresponding cleaning operator $o$ is executed, producing $\mathbf{X}^{(l+1)}$ and appending $o$ to $P$ (Line 9).

After applying $o$, $E$ re-evaluates $\mathbf{X}^{(l+1)}$ and $M_{lite}$ computes the dense lightweight model reward $\mathcal{R}_{lite}^{dense}$ (Lines 10--11). The low-level reward updates $\theta_L$, and the high-level dense reward updates $\theta_H$ (Lines 12--13). The step counter $l$ is then incremented (Line 14).
When the inner loop ends (either triggered by \textnormal{F} or reaching $\mathcal{L}_{\max}$), the full pipeline is validated by $M_{complex}$. The sparse high-level reward $\mathcal{R}_{high}^{sparse}$ is computed from the performance gain over the raw data and is used to update $\theta_H$ (Line 16). Finally, the trained agents $Agent_H$ and $Agent_L$ are returned (Line 18).

% Once the training phase in Algorithm~\ref{alg:hierarchical_cleaning} is complete, the system enters the inference phase. The optimized agents greedily execute their policies on any given dirty dataset $\mathbf{X}$ until $Agent_H$ outputs \textnormal{FINISH}. This deterministic execution skips the model updating steps, directly producing the final pipeline $P^*$ and the cleaned dataset $\mathbf{X}_{cleaned}$ for downstream tasks.

\subsubsection{Accelerated Convergence}
%We discuss two heuristic mechanisms to improve learning efficiency and ensure policy convergence. %, two heuristic mechanisms are proposed as follows. 
A natural question arises regarding the convergence of our algorithm. In hierarchical decision-making for data cleaning, the high-level agent typically explores different actions (e.g., missing value imputation, outlier correction, and constraint violation repair) to find an optimal sequence. However, when the current data contains missing values, most cleaning operations (such as outlier detection) are prone to error. This makes searching over a wide action space inefficient and potentially harmful. An intuitive strategy is to prioritize imputation immediately upon detecting missing values, effectively eliminating other choices. This not only simplifies the agent’s decision, but also ensures the dataset is structurally complete before executing subsequent cleaning. The following lemma formalizes this dependency-aware pruning.

% \begin{lemma}[Dependency-aware Search Pruning]
%     The high-level agent's action space $\mathcal{A}^H$ is exclusively pruned to $M$ upon the identification of null values in the current data.
% \label{lemma1} 
% \end{lemma}
% The proof is straightforward. If null values are identified in the current data, it means the data contains missing values. As we adopt an imputation-first priority on the high-level agent, the imputation is adopted directly without searching in the high-level agent's action space. This ensures a consistent data foundation, effectively narrowing the search space to valid trajectories.

\begin{lemma}[Dependency-aware Search Pruning]
    If missing values are detected in the current data, the valid high-level action space is temporarily restricted to 
    $\mathcal{A}^H_{\text{valid}}=\{\text{M}\}$, i.e., the agent must first perform missing-value repair.
\label{lemma1} 
\end{lemma}
When null values are present, non-imputation operations such as outlier detection or constraint checking may be unreliable. Therefore, we temporarily mask all actions except $\text{M}$ and restore the full action space after missing values are repaired, thereby avoiding invalid cleaning trajectories.

In addition, we adopt a Stagnation-triggered Penalty Strategy ($\mathcal{R}_{penalty}$)  to prevent infinite loops and stagnation. Specifically, if the agent executes the exact same action pair $(a^H, a^L)$ for three consecutive steps without any substantive performance improvement, a massive penalty is directly added to $\mathcal{R}_{high}^{dense}$ (Line 13). Formally, $\mathcal{R}_{penalty}$ is set to $5.0$ when this stagnation condition is triggered, and remains $0$ otherwise. This rigorous constraint deliberately outweighs the incremental gains from other reward components, forcibly breaking behavioral deadlocks and prompting the exploration of alternative operators.

% \begin{algorithm}[tb]
% \small \DontPrintSemicolon \SetAlgoLined
% \caption{Data Cleaning Using Trained Hierarchical Agents}
% \label{alg:cleaning_inference}

% \KwIn{Dirty dataset $\mathbf{X}_{dirty}$; trained high-level agent $Agent_H(\theta_H)$; trained low-level agent $Agent_L(\theta_L)$; operator library $\mathcal{B}$; quality evaluator $E$}
% \KwOut{Cleaned dataset $\mathbf{X}_{cleaned}$; final pipeline $P^*$}

% $(\mathbf{X}^{(0)}, P, l) \leftarrow (\mathbf{X}_{dirty}, \emptyset, 0)$\;

% \While{True}{
%   $\mathbf{Rates}^{(l)} \leftarrow E(\mathbf{X}^{(l)})$\;
  
%   $a^H \leftarrow Agent_H(\textnormal{ExtractHighState}(\mathbf{X}^{(l)}, \mathbf{Rates}^{(l)}))$\;
%   \lIf{$a^H = \textnormal{F}$}{\textbf{break}}
  
%   $a^L \leftarrow Agent_L(\textnormal{ExtractLowState}(\mathbf{X}^{(l)}, a^H))$\;
  
%   $o \leftarrow \mathcal{B}(a^H, a^L)$\;
%   $(\mathbf{X}^{(l+1)}, P) \leftarrow (o(\mathbf{X}^{(l)}), P \oplus o)$\;
  
%   $l \leftarrow l + 1$\;
% }

% $\mathbf{X}_{cleaned} \leftarrow \mathbf{X}^{(l)}$, $P^* \leftarrow P$\;
% \Return{$\mathbf{X}_{cleaned}$, $P^*$}\;
% \end{algorithm}

\begin{algorithm}[tb]
\small \DontPrintSemicolon \SetAlgoLined
\caption{Data Cleaning Using Trained Hierarchical Agents}
\label{alg:cleaning_inference}

\KwIn{Dirty dataset $\mathbf{X}_{dirty}$; frozen high-level Q-table $Q_H$; frozen low-level Q-tables $Q_L$; operator library $\mathcal{B}$; quality evaluator $E$; maximum pipeline length $\mathcal{L}_{\max}$}
\KwOut{Cleaned dataset $\mathbf{X}_{cleaned}$; final pipeline $P^*$}

$(\mathbf{X}^{(0)}, P, l) \leftarrow (\mathbf{X}_{dirty}, \emptyset, 0)$\;

\While{$l < \mathcal{L}_{\max}$}{
  $\mathbf{Rates}^{(l)} \leftarrow E(\mathbf{X}^{(l)})$\;

  $s^H \leftarrow \textnormal{ExtractHighState}(\mathbf{X}^{(l)}, \mathbf{Rates}^{(l)})$\;
  $a^H \leftarrow \textnormal{GreedySelect}_H(s^H, Q_H)$

  % $a^H \leftarrow \textnormal{GreedySelect}_H(Q_H, ExtractHighState(\mathbf{X}^{(l)}, \mathbf{Rates}^{(l)}))$  

  \lIf{$a^H = \textnormal{F}$}{\textbf{break}}

  $s^L \leftarrow \textnormal{ExtractLowState}(\mathbf{X}^{(l)}, a^H)$\;
  $a^L \leftarrow \textnormal{GreedySelect}_L(s^L, Q_L)$
  
  % $a^L \leftarrow \textnormal{GreedySelect}_L(Q_L, ExtractLowState(\mathbf{X}^{(l)}, a^H))$
  
  $o \leftarrow \mathcal{B}(a^H, a^L)$\;
  $(\mathbf{X}^{(l+1)}, P) \leftarrow (o(\mathbf{X}^{(l)}), P \oplus o)$\;
  
  $l \leftarrow l + 1$\;
}

$\mathbf{X}_{cleaned} \leftarrow \mathbf{X}^{(l)}$, $P^* \leftarrow P$\;
\Return{$\mathbf{X}_{cleaned}$, $P^*$}\;
\end{algorithm}

\subsection{Cleaning Inference}

Once training completes and given a new dirty dataset $\mathbf{X}_{dirty}$, the system starts the inference phase detailed in Algorithm~\ref{alg:cleaning_inference}. At each step, the quality evaluator first computes the current quality indicators and extracts the high-level state (Lines~3--4). The high-level agent then selects the quality issue with the maximum Q-value under the current state (Line~5). If the termination action is not selected, the low-level state is constructed and the low-level agent greedily selects the operator with the highest learned value from the corresponding sub-library in $\mathcal{B}$ (Lines~7--9).

The selected operator is applied immediately, incrementally transforming the data and updating the pipeline (Line~10). This process continues until the high-level agent selects the termination action ($a^H=\mathrm{F}$) or the maximum pipeline budget $\mathcal{L}_{\max}$ is reached (Lines~2 and~6). The final cleaned dataset $\mathbf{X}_{cleaned}$ and pipeline $P^*$ are then returned (Lines~13--14). Since the learned Q-values are frozen and all actions are selected greedily during inference, given the same input data, trained Q-tables, quality evaluator, and operator library, \sys produces the same explicit and auditable cleaning pipeline. Exploration and Q-table updates are performed only during training.

\subsection{Complexity Analysis}

%Let $\mathcal{L}_{\max}$ be the pipeline budget, $K$ the operator-pool size, and $K_{\max}$ the maximum number of operators within any single sub-library.
%At each cleaning iteration $l$, the high-level agent chooses an issue type from $\mathcal{A}^H$, and the low-level agent then selects an operator from the corresponding sub-library.  The decision cost per step is bounded by $\mathcal{O}(|\mathcal{A}^H| + K_{\max})$. Since both the number of issue types $|\mathcal{A}^H|$ and the maximum sub-library size $K_{\max}$ are upper-bounded by $K$, the cost per step simplifies to $\mathcal{O}(K)$. Thus, the complexity of one episode \mkh{?} \textcolor{blue}{(What makes you confused? Check the Algorithm 1)}of cleaning pipeline generation is: $\mathcal{O}(\mathcal{L}_{\max}K)$. Given that the number of training episodes $N$ is a constant, the total training cost remains asymptotically: $\mathcal{O}(\mathcal{L}_{\max}K)$. \mkh{This complexity is about pipeline generation? What about the cost of pipeline selection? Also, what about the cost of the whole cleaning process?} \textcolor{blue}{(The number of training episodes $N$ in Algorithm 1 indicates the whole training process of the agents, and then we use the agents to directly generate the cleaning pipeline! The complexity is the whole process!)}

We analyze the computational complexity of both training and inference. Let $\mathcal{L}_{\max}$ be the maximum pipeline length (budget), $K = |\mathcal{B}|$ the total cleaning operator pool size, $K_{\max}$ the maximum size of any single sub-library (i.e., $\max_{a^H \in \mathcal{A}^H} |\mathcal{B}_{a^H}|$), and $N$ the number of training episodes. We assume $|\mathcal{A}^H| \leq K$ and $K_{\max} \leq K$.

%\subsubsection{Per-Step Decision Cost}
At each cleaning iteration $l$, the high-level agent selects an action from $\mathcal{A}^H$ (cost $\mathcal{O}(|\mathcal{A}^H|)$), and the low-level agent selects an operator from the corresponding sub-library (cost $\mathcal{O}(K_{\max})$). Thus, the decision cost per step is:
\begin{equation}
    \mathcal{O}(|\mathcal{A}^H| + K_{\max}) = \mathcal{O}(K).
\end{equation}

%\subsubsection{Training Complexity (Algorithm~\ref{alg:hierarchical_cleaning})}
For each training episode, the inner loop executes at most $\mathcal{L}_{\max}$ steps. In each step, besides the decision cost $\mathcal{O}(K)$, the algorithm (i) evaluates quality issue rates via $E(\mathbf{X}^{(l)})$: typically $\mathcal{O}(d)$ where $d$ is the dataset size; (ii) computes dense reward $\mathcal{R}_{lite}^{dense}$ using lightweight model $M_{lite}$: cost $\mathcal{O}(d \cdot h_{lite})$ where $h_{lite}$ is the model's hidden dimension; and
(iii) updates policy parameters $\theta_L$ and $\theta_H$: cost proportional to model size, typically $\mathcal{O}(|\theta_L| + |\theta_H|)$.

At the end of each episode, the sparse reward requires evaluating the complex model $M_{complex}$ on the cleaned data $\mathbf{X}^{(l)}$, with cost $\mathcal{O}(d \cdot h_{complex})$, where $h_{complex} \gg h_{lite}$.

Let $C_E(d)$ denote the cost of quality evaluation, $C_{op}(d)$ the cost of applying a cleaning operator, $C_{lite}(d)$ the cost of the lightweight model forward pass, $C_{complex}(d)$ the cost of the complex model forward pass, and $C_{\theta}$ the cost of policy updates (independent of $d$). The total training cost is:
\begin{equation}
\resizebox{\linewidth}{!}{$
\mathcal{O}\!\left(N \cdot \mathcal{L}_{\max} \cdot \left(K + C_E(d) + C_{op}(d) + C_{lite}(d) + C_{\theta}\right) + N \cdot C_{complex}(d)\right)
$}
\end{equation}

% During inference, no model updates or reward computations occur.
During inference, no policy updates or downstream reward evaluations are performed. At each step, the agents perform action selection $\mathcal{O}(K)$ and apply one cleaning operator (e.g., $\mathcal{O}(d)$ for operations like imputation). Quality evaluation $E$ is also required at each step to extract the state. Hence, to sum up, the inference cost for cleaning is:
\begin{equation}
\mathcal{O}\!\left(\mathcal{L}_{\max} \cdot \left(K + C_E(d) + C_{op}(d)\right)\right).
\end{equation}

Thus, under fixed per-step evaluation and operator costs, the computational cost grows linearly with the pipeline budget \(\mathcal{L}_{\max}\), and the training cost additionally grows linearly with the size of the candidate operator pool $K$.

% Thus, both the training and inference phases of our method scale linearly with respect to the pipeline budget $\mathcal{L}_{\max}$ and the number of cleaning operations $K$.

Compared with an exhaustive enumeration baseline, \sys avoids the exponential search over cleaning pipelines. Specifically, such a baseline enumerates all length-\(\mathcal{L}_{\max}\) pipelines from an operator pool of size \(K\), resulting in \(K^{\mathcal{L}_{\max}}\) candidate pipelines. For each candidate pipeline, the baseline sequentially executes \(\mathcal{L}_{\max}\) cleaning operators, and then evaluates the resulting data quality and downstream utility once. This leads to an intractable overall cost of
$
\mathcal{O}\!\left(
K^{\mathcal{L}_{\max}}
\cdot
\left(
\mathcal{L}_{\max} \cdot C_{op}(d) + C_E(d) + C_{complex}(d)
\right)
\right)
$.

% Therefore, compared with the exhaustive enumeration baseline, which requires executing and evaluating every possible pipeline on the downstream task with an intractable complexity of:
% \[
% \mathcal{O}\!\left(K^{\mathcal{L}_{\max}} \cdot \left(C_{op}(d) + C_{complex}(d)\right)\right),
% \]
% our method avoids exponential explosion and scales strictly linearly with respect to both the pipeline budget $\mathcal{L}_{\max}$ and the number of training episodes $N$, demonstrating superior scalability.
\section{Experimental Evaluation}
\label{sec:eval}

In this section, we validate our technical contributions through a series of experiments, divided into two main sets. The first evaluates \sys in terms of upstream and downstream performance as well as runtime. The second examines how individual components of \sys impact its effectiveness and efficiency.

%\mkh{give a summary of the experiments} \textcolor{blue}{These results demonstrate the effectiveness and practical applicability of \sys across diverse real-world time-series scenarios.} 

\subsection{Experimental Setup}

\begin{table}[tb]
  \caption{Dataset Statistics}
  \vspace{-3mm}
  \label{tab:datasets}
  \centering
  \setlength{\tabcolsep}{2pt} % 默认6pt，可以调小
  \begin{tabular}{lcccccl}
    \toprule
    Dataset & \#Categories & Length & \#Samples & \#Attrs & Task \\
    \midrule
    ETTh1 & 1 & 17,420 & 1 & 7 & Pred. \\
    IDF\_OilTemp  & 1 & 1,024 & 1 & 4 & Pred. \\
    Libras  & 15 & 45 & 360 & 2 & Clf./Clu. \\
    Handwriting  & 26 & 152 & 1,000 & 3 & Clf./Clu. \\
    \bottomrule
  \end{tabular}
%  \vspace{-2mm}
\end{table}

\subsubsection{Datasets} 

We evaluate \sys on four widely used real-world datasets from diverse domains, as summarized in Table~\ref{tab:datasets}. 

\begin{itemize}[topsep=0pt, partopsep=0pt]
\item ETTh1~\cite{DBLP:conf/aaai/ZhouZPZLXZ21}: a power dataset collected from power transformers across two regions, capturing two years of hourly operational data with seven attributes, including the target oil temperature and six external power load attributes.

\item IDF\_OilTemp~\cite{DBLP:journals/pvldb/DingSWWY24}: a heavy-equipment monitoring dataset with hourly sequences, where each sequence contains 1,024 time steps and four continuous attributes representing the main oil temperature and correlated mechanical states.

\item Libras~\cite{DBLP:journals/corr/abs-1811-00075}: a Brazilian Sign Language dataset with 360 samples across 15 classes, where each sequence contains 45 time steps recording the spatial coordinates of the hand centroid.

\item Handwriting~\cite{DBLP:journals/corr/abs-1811-00075}: a smartwatch-based dataset with 1,000 samples across 26 classes, capturing alphabet-writing motions, where each sequence contains 152 time steps of triaxial real-time acceleration signals.
\end{itemize}

% \begin{itemize}[topsep=0pt, partopsep=0pt]
% \item ETTh1~\cite{DBLP:conf/aaai/ZhouZPZLXZ21}: a power dataset collected from power transformers across two regions, capturing two years of hourly operational data with seven attributes (target oil temperature and six external power loads). 

% \item IDF\_OilTemp~\cite{DBLP:journals/pvldb/DingSWWY24}: a heavy-equipment monitoring dataset with hourly sequences (length 1,024) and four continuous attributes representing the main oil temperature and correlated mechanical states.

% \item Libras~\cite{DBLP:journals/corr/abs-1811-00075}: a Brazilian Sign Language dataset with 360 samples across 15 classes, where each length-45 sequence records the spatial coordinates of the hand centroid. 

% \item Handwriting~\cite{DBLP:journals/corr/abs-1811-00075}: a smartwatch-based dataset with 1,000 samples across 26 classes, capturing alphabet-writing motions, where each sequence of length 152 consists of triaxial real-time acceleration signals. 
% \end{itemize}

ETTh1 and IDF\_OilTemp are long-sequence datasets without class labels used for forecasting (Pred.), whereas Libras and Handwriting are labeled multi-class datasets used for classification (Clf.) and clustering (Clu.) tasks.

Real-world dirty MTS datasets often lack clean ground truth, making it difficult to quantitatively evaluate cleaning effectiveness. They also rarely provide complete annotations of co-existing quality issues, which limits systematic evaluation across different issue types. Therefore, following MTSClean~\cite{DBLP:journals/pvldb/DingSWWY24}, we inject controlled corruptions and additive Gaussian white noise into real-world MTS datasets to obtain dirty-clean pairs, covering duplicate values, missing values, single-point and segment outliers, and constraint violations on randomly selected attributes.

% To simulate real-world data quality degradation, we inject discrete corruptions and additive Gaussian white noise following the MTSClean contamination strategy~\cite{DBLP:journals/pvldb/DingSWWY24}.
% Specifically, we inject four distinct types of data quality issues, including duplicate values, missing values, single-point and segment outliers, and constraint violations, into randomly selected attributes. 

\subsubsection{Baselines}

We compare \sys against four widely used cleaning methods: 

\begin{itemize}[topsep=0pt, partopsep=0pt]
\item EDITOR~\cite{li2026editor}: A recent MTS cleaning method that handles multi-granularity issues (points, subsequences, cross-attribute) via a multi-resolution pipeline for detection, localization, and repair, while mitigating over-cleaning.

\item  Clean4TSDB~\cite{DBLP:journals/pvldb/DingSWYWW24}: An integrated system combining constraint mining and error profiling to detect violations of physical and statistical patterns, and repair them via temporal and multivariate dependencies.

\item DiffPrep~\cite{DBLP:journals/pacmmod/LiCC023}: The state-of-the-art tabular pipeline generation method that maps discrete preprocessing choices into a differentiable space, enabling joint optimization of pipelines and downstream model parameters. 
Designed for standard tabular data, we adapt it to time-series cleaning by replacing general preprocessing operators with time-series-specific cleaning operators. The adaptation works only for single-series datasets.

\item Sampling: It adopts a brute-force pipeline generation strategy by randomly sampling up to $\mathcal{L}_{\max}$ operators from the predefined cleaning-operator pool and permuting them into a random execution order to clean the dataset.

% We compare \sys against three recent automated cleaning systems that represent diverse technical paradigms in data preparation and time-series cleaning: \mkh{make sure the rationale for specifically choosing these techniques is clear! Why not all the techniques mentioned in the intro or related work sections?}
\end{itemize}

\subsubsection{Evaluation Metrics}

We evaluate our method from two complementary perspectives. For upstream data cleaning, we use three metrics: F1-score \cite{DBLP:journals/pvldb/DingSWWY24} for quality-issue detection; Normalized Mean Squared Error (NMSE) \cite{DBLP:conf/nips/CaoWLZLL18} and Relative Repair Accuracy (RRA) \cite{DBLP:journals/pvldb/DingSWWY24} for data cleaning. 

To evaluate downstream performance, we adopt task-specific evaluation metrics. For forecasting, we compute the average of exponentially scaled NRMSE ($e^{-\text{NRMSE}}$) and the normalized Correlation Coefficient (CC), capturing both prediction accuracy and alignment of temporal trends. For classification, we average the Macro F1-score and ROC AUC, reflecting performance across balanced classes as well as threshold-independent discriminative power. For clustering, we use the average of the normalized Silhouette Score and the inversely scaled Davies–Bouldin Index, which together quantify intra-cluster cohesion and inter-cluster separation. All downstream metrics are normalized to the range $[0,1]$, with higher values indicating better performance, ensuring comparability across different tasks. Finally, $\Delta$Perf represents the relative improvement in downstream performance.

\subsubsection{Downstream Models}
In the Dual-Stage Reward Mechanism, we employ DLinear~\cite{zeng2023transformers} for forecasting, MiniRocketClassifier~\cite{DBLP:conf/kdd/DempsterSW21} for classification, and Catch22Clusterer~\cite{DBLP:journals/datamine/LubbaSKSFJ19} for clustering as lightweight proxy models. These models are widely used baselines in their respective tasks and are computationally efficient. For final end-to-end downstream utility, we evaluate each task using a corresponding complex target model: LSTMForecast~\cite{hochreiter1997long}, InceptionTimeClassifier~\cite{DBLP:journals/datamine/FawazLFPSWWIMP20}, and AEDCNNClusterer~\cite{DBLP:journals/corr/abs-1803-01271}, respectively. These target models provide strong representational capacity for forecasting, classification, and clustering.

% This two-model setup indicates that the cleaning effectiveness of \sys is not overly dependent on a specific downstream model, and remains stable across lightweight and complex model, demonstrating robustness to downstream model choice.

\subsubsection{Implementation}
All experiments are conducted on a Linux workstation (Ubuntu 24.04.1 LTS) equipped with an Intel Core i9-10900K CPU, 128 GB of RAM, and an NVIDIA RTX 3090 GPU (24 GB VRAM). The source code can be accessed at \url{https://anonymous.4open.science/r/AegisTS-4131/}.

\begin{table*}[tb]
  \centering
  \small 
  \renewcommand{\arraystretch}{0.8} 
  
  \caption{Overall comparison on all datasets across different downstream tasks.} \vspace{-3mm}
  \label{tab:comprehensive_evaluation}
  \small
  %\resizebox{\linewidth}{!}{
  \begin{tabular}{c c@{\quad} ccc@{\quad} cc@{\qquad} ccc@{\quad} cc} 
    \toprule[1pt]
    
    % ================= 第一部分 =================
    \multirow{3}{*}{Task} & \multirow{3}{*}{Method} & \multicolumn{5}{c}{ETTh1} & \multicolumn{5}{c}{IDF\_OilTemp} \\
    \cmidrule(lr){3-7} \cmidrule(lr){8-12}
    & & \multicolumn{3}{c}{Upstream} & \multicolumn{2}{c}{Downstream} 
    & \multicolumn{3}{c}{Upstream} & \multicolumn{2}{c}{Downstream} \\
    \cmidrule(lr){3-5} \cmidrule(lr){6-7} \cmidrule(lr){8-10} \cmidrule(lr){11-12}
    & & F1$\uparrow$ & NMSE$\downarrow$ & RRA$\uparrow$ & Perf$\uparrow$ & $\Delta$Perf$\uparrow$ 
    & F1$\uparrow$ & NMSE$\downarrow$ & RRA$\uparrow$ & Perf$\uparrow$ & $\Delta$Perf$\uparrow$ \\
    \cmidrule{1-12}
    
    \multirow{5}{*}{\rotatebox{90}{Forecasting}} 
    & Sampling & 0.2419 & 0.0837 & 0.2245 & 0.8790 & -0.0205 
    & 0.3733 & 0.0095 & 0.9905 & 0.9842 & +0.1635 \\
    
    & DiffPrep & 0.6680 & 0.0041 & 0.7915 & 0.9334 & +0.0339 
    & 0.2991 & 0.0113 & 0.9909 & 0.9767 & +0.1560 \\
    
    & EDITOR & 0.2044 & 0.1112 & 0.2557 & 0.9100 & +0.0105 
    & 0.2215 & 0.0059 & 0.9647 & 0.9659 & +0.1452 \\
    
    & Clean4TSDB & 0.2393 & 0.0050 & 0.3522 & 0.9243 & +0.0248 
    & 0.0561 & \textbf{0.0058} & 0.0895 & 0.8355 & +0.0148 \\
    
    \cmidrule(lr){2-12}
    & \textbf{\sys} & \textbf{0.6758} & \textbf{0.0040} & \textbf{0.7966} & \textbf{0.9437} & \textbf{+0.0442} 
    & \textbf{0.5546} & 0.0082 & \textbf{0.9924} & \textbf{0.9873} & \textbf{+0.1666} \\
    
    \midrule[1pt]
    
    % ================= 第二部分 =================
    \multirow{3}{*}{Task} & \multirow{3}{*}{Method} & \multicolumn{5}{c}{Libras} & \multicolumn{5}{c}{Handwriting} \\
    \cmidrule(lr){3-7} \cmidrule(lr){8-12}
    & & \multicolumn{3}{c}{Upstream} & \multicolumn{2}{c}{Downstream} 
    & \multicolumn{3}{c}{Upstream} & \multicolumn{2}{c}{Downstream} \\
    \cmidrule(lr){3-5} \cmidrule(lr){6-7} \cmidrule(lr){8-10} \cmidrule(lr){11-12}
    & & F1$\uparrow$ & NMSE$\downarrow$ & RRA$\uparrow$ & Perf$\uparrow$ & $\Delta$Perf$\uparrow$ 
    & F1$\uparrow$ & NMSE$\downarrow$ & RRA$\uparrow$ & Perf$\uparrow$ & $\Delta$Perf$\uparrow$ \\
    \cmidrule{1-12}
    
    \multirow{4}{*}{\rotatebox{90}{Classif.}} 
    & Sampling & 0.1967 & 0.0018 & 0.8056 & 0.8216 & +0.0043 
    & 0.1871 & 0.0035 & 0.1574 & 0.4182 & -0.1078 \\
    
    & EDITOR & \textbf{0.6547} & 0.0162 & 0.5132 & 0.8191 & +0.0018 
    & 0.3779 & 0.0010 & \textbf{0.4755} & 0.5281 & +0.0021 \\
    
    & Clean4TSDB & 0.0197 & 0.0019 & 0.0372 & 0.9046 & +0.0873 
    & 0.1373 & 0.0032 & 0.1170 & 0.4634 & -0.0626 \\
    
    \cmidrule(lr){2-12}
    & \textbf{\sys} & 0.4898 & \textbf{0.0010} & \textbf{0.8374} & \textbf{0.9506} & \textbf{+0.1333} 
    & \textbf{0.5450} & \textbf{0.0005} & 0.2034 & \textbf{0.5920} & \textbf{+0.0660}  \\
    
    \midrule[1pt]
    
    % ================= 第三部分 =================
    \multirow{3}{*}{Task} & \multirow{3}{*}{Method} & \multicolumn{5}{c}{Libras} & \multicolumn{5}{c}{Handwriting} \\
    \cmidrule(lr){3-7} \cmidrule(lr){8-12}
    & & \multicolumn{3}{c}{Upstream} & \multicolumn{2}{c}{Downstream} 
    & \multicolumn{3}{c}{Upstream} & \multicolumn{2}{c}{Downstream} \\
    \cmidrule(lr){3-5} \cmidrule(lr){6-7} \cmidrule(lr){8-10} \cmidrule(lr){11-12}
    & & F1$\uparrow$ & NMSE$\downarrow$ & RRA$\uparrow$ & Perf$\uparrow$ & $\Delta$Perf$\uparrow$ 
    & F1$\uparrow$ & NMSE$\downarrow$ & RRA$\uparrow$ & Perf$\uparrow$ & $\Delta$Perf$\uparrow$ \\
    \cmidrule{1-12}
    
    \multirow{4}{*}{\rotatebox{90}{Clustering}} 
    & Sampling & 0.1759 & 0.0092 & 0.5157 & 0.6704 & +0.0198 
    & 0.2031 & 0.0032 & 0.1399 & 0.6009 & +0.0605 \\
    
    & EDITOR & \textbf{0.6546} & 0.0215 & 0.4197 & 0.6760 & +0.0254 
    & 0.3464 & \textbf{0.0011} & \textbf{0.4471} & 0.5674 & +0.0270 \\
    
    & Clean4TSDB & 0.0145 & 0.0019 & 0.0284 & 0.6898 & +0.0392 
    & 0.1346 & 0.0018 & 0.1485 & 0.5588 & +0.0184 \\
    
    \cmidrule(lr){2-12}
    & \textbf{\sys} & 0.4731 & \textbf{0.0018} & \textbf{0.8030} & \textbf{0.6930} & \textbf{+0.0424} 
    & \textbf{0.4033} & 0.0016 & 0.2716 & \textbf{0.6537} & \textbf{+0.1133} \\
    
    \bottomrule[1pt]
  \end{tabular}
%  }
\end{table*}

\begin{table}[tb]
  \renewcommand{\arraystretch}{0.8} 
  \caption{Generalization evaluation across different datasets} \vspace{-3mm}
  %\vspace{-4mm}
  \label{tab:generalization_evaluation}
  \centering
  \begin{tabular}{lccc}
    \toprule
    Task & Dataset & NMSE$ \downarrow$ & $\Delta$Perf$ \uparrow$ \\
    \midrule
    \multirow{2}{*}{Forecasting} & IDF\_OilTemp$\rightarrow$ETTh1 & 0.0050 & +0.0290 \\
                                 & ETTh1$\rightarrow$IDF\_OilTemp & 0.0077 & +0.1589 \\
    \midrule
    \multirow{2}{*}{Classification} & Handwriting$\rightarrow$Libras & 0.0044 & -0.0034 \\
                                    & Libras$\rightarrow$Handwriting & 0.0012 & +0.0395 \\
    \midrule
    \multirow{2}{*}{Clustering} & Handwriting$\rightarrow$Libras & 0.0085 & +0.0335 \\
                                & Libras$\rightarrow$Handwriting & 0.0026 & +0.0403 \\
    \bottomrule
  \end{tabular}
 % \vspace{-3mm}
\end{table}

% \begin{table*}[!t]
%   \caption{Time cost of different cleaning methods}
%  % \vspace{-4mm}
%   \label{tab:efficiency_evaluation}
%   \centering
%   \resizebox{\textwidth}{!}{%
%   \begin{tabular}{lcccccc}
%     \toprule
%     Method & ETTh1 (Pred.) & IDF\_OilTemp (Pred.) & Libras (Clf.) & Handwriting (Clf.) & Libras (Clu.) & Handwriting (Clu.) \\
%     \midrule
    
%     Sampling & 252 s & 1071 s & 864 s & 1497 s & 892 s & 916 s \\
%     DiffPrep & 604 s & 174 s & -- & -- & -- & -- \\
%     EDITOR & 1345 s & 995 s & 48 s & 501 s & 48 s & 484 s \\
%     Clean4TSDB & 848 s & 770 s & 164 s & 6236 s & 151 s & 6317 s \\
%     \sys-Transfer & 738 s & 109 s & 3742 s & 6236 s & 3278 s & 5068 s \\
%     \midrule

%     Brute-Force Search & > 3 days & > 3 days & > 3 days & > 3 days & > 3 days & > 3 days \\
%     Single Agent & 4312 s & 1208 s & 32930 s & 60103 s & 30606 s & 58302 s \\
%     \midrule
%     \sys & 2016 s & 627 s & 17617 s & 21674 s & 16193 s & 20967 s \\
%     \bottomrule
%   \end{tabular}%
%   }
% \end{table*}

%\vspace{-2mm}
\subsection{Cleaning Performance}

Table~\ref{tab:comprehensive_evaluation} presents a comprehensive comparison of all methods across six dataset-task configurations. We analyze the experimental results from the perspectives of both upstream and downstream metrics.

\softbsubsec{Upstream Performance} The results show that \sys consistently outperforms all baselines across most upstream performance settings. We further break down the results by dataset type.
%analyze the upstream metrics across different tasks as follows.

%\textbf{1) Forecasting:} 
In the ETTh1 forecasting dataset, \sys dominates across all upstream metrics, achieving a 96.4\% $NMSE$ reduction compared with EDITOR. This is because \sys leverages a robust constraint miner to capture physical and cross-attribute dependencies from dirty data, preventing error accumulation. In addition, we observe the strategic trade-off on the IDF\_OilTemp dataset: while Clean4TSDB achieves a marginally lower $NMSE$, \sys achieves a significantly higher $F1$ score and a near-optimal $RRA$. This is attributed to the fact that the cleaning strategy of \sys avoids the over-smoothing driven solely by minimizing numerical deviations, and instead prioritizes the recovery of meaningful temporal patterns. 

% \textbf{2) Classification:} While EDITOR achieves higher scores on specific upstream metrics, such as $F1$ on Libras and $RRA$ on Handwriting—these improvements do not translate into meaningful downstream performance gains (+0.0018 and +0.0021). This indicates that its cleaning process lacks practical effectiveness in preserving the discriminative attributes and class boundaries required for classification. In contrast, \sys prioritizes the restoration of category-specific patterns, resulting in substantial downstream improvements across both datasets. Overall, \sys ensures that the cleaned data remains highly informative for classification by avoiding overemphasis on upstream metrics and focusing on discriminative attribute preservation.

%\textbf{2) Classification:} 
In the Libras classification dataset,  \sys achieves higher cleaning quality (NMSE: 0.0010, RRA: 0.8374), while EDITOR attains higher F1. In the Handwriting dataset, \sys obtains a lower RRA than EDITOR. This is because \sys emphasizes structure-preserving repairs, jointly using constraint modeling to avoid overfitting to local anomalies and better preserve global patterns. In contrast, EDITOR relies on multi-resolution local detection, tending to optimize specific metrics while potentially compromising global structural consistency, which is critical for downstream analytical tasks like classification. Note that DiffPrep's results are not reported, as it cannot handle multi-series datasets.

%\textbf{3) Clustering:} 
The clustering results show clear dataset-dependent trends. On Libras (low-frequency, smooth trajectories), \sys achieves the best cleaning quality (NMSE: 0.0018, RRA: 0.8030), while EDITOR attains a higher F1 (0.6546) due to its high-sensitivity detection. On Handwriting (high-frequency), \sys shows stronger anomaly identification (F1: 0.4033), whereas EDITOR's smoothing (low-pass filtering) reduces point-wise error. In summary, \sys prioritizes structural fidelity, preserving meaningful dynamics for anomaly detection, while EDITOR favors accuracy through smoothing.

\softbsubsec{Downstream Performance} \sys consistently outperforms the baselines, achieving the highest absolute performance ($Perf$) and performance gains ($\Delta Perf$) across all evaluated scenarios. This significant advantage stems from its robust capabilities in three specific downstream tasks:

% \textbf{1) Forecasting:} On both the ETTh1 and IDF\_OilTemp datasets, \sys achieves the highest downstream gains because of its effectiveness in restoring complex temporal dynamics and preserving forecasting-relevant patterns.

%\textbf{1) Forecasting:}
In forecasting, \sys achieves the highest downstream gains on ETTh1 ($\Delta$Perf: +0.0442) and IDF\_OilTemp (+0.1666) by effectively restoring complex temporal dynamics. While baselines like Clean4TSDB minimize numerical errors at the risk of over-smoothing critical trends, \sys leverages its dual-stage reward mechanism to explicitly balance structural data fidelity with downstream utility. This ensures the preservation of meaningful temporal patterns, yielding a substantial performance boost.

%\textbf{2) Classification:}
In classification, \sys consistently outperforms all baselines. In the highly sensitive Handwriting dataset, baselines such as Clean4TSDB degrade model accuracy, yielding a negative downstream impact ($\Delta Perf$: -0.0626). In contrast, \sys safely cleans the data to deliver a robust positive gain (+0.0660), achieving a 27.75\% improvement in absolute performance ($Perf$) over Clean4TSDB. This highlights a limitation of task-agnostic cleaning methods such as Clean4TSDB, where statistical smoothing can erase subtle stroke dynamics. In contrast, \sys uses dual-stage rewards to penalize repairs that damage class boundaries, preserving multi-class motion patterns while removing corruptions.

%\textbf{3) Clustering:} 
In clustering,  the results also show the robustness of \sys in unsupervised settings by preserving intrinsic spatial structures. The downstream gain on Handwriting ($\Delta$Perf: +0.1133) significantly exceeds that on Libras (+0.0424). This stems from Handwriting's higher complexity (length-152 triaxial signals vs. length-45 spatial coordinates) and higher error rate (26.48\% vs. 22.96\%). Under such severe corruption, conventional methods tend to over-smooth and erase subtle stroke dynamics. \sys overcomes this via dual-stage rewards that penalize structural damage, successfully preserving intricate motion patterns and yielding larger improvements on the more challenging dataset.

\begin{figure*}[tb]
  \centering
  \includegraphics[width=0.8\textwidth,height=\linewidth,keepaspectratio]{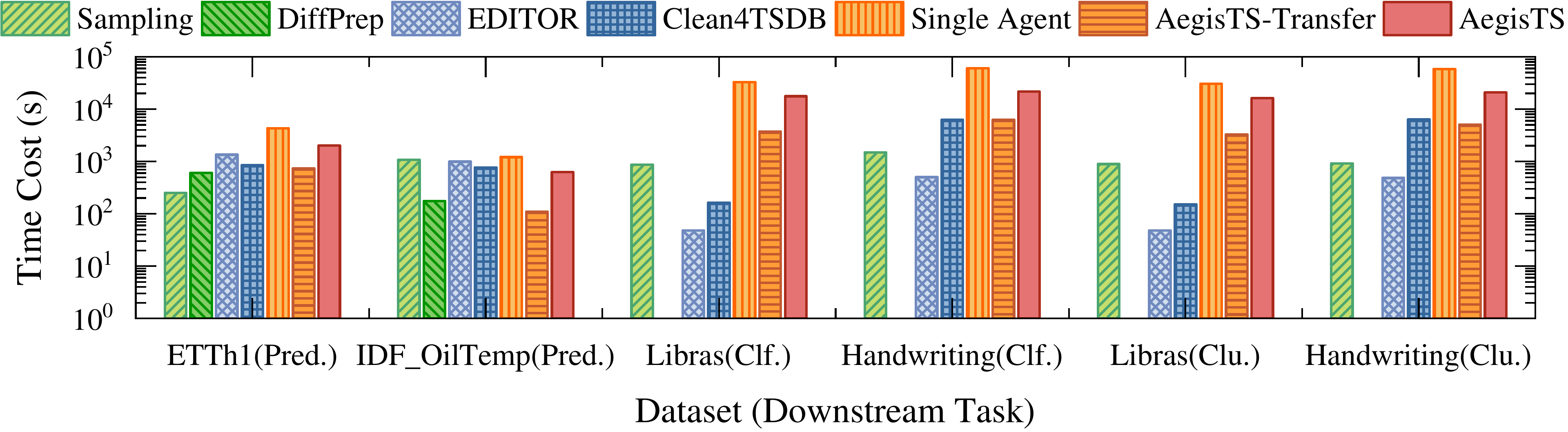} \vspace{-3mm}
  \caption{Time cost of different cleaning methods. Brute-Force Search is omitted for readability because it takes more than three days on all datasets.} 
  \Description{Time and Performance.}
  \label{fig:time}
\end{figure*}

\subsection{Generalization Evaluation}
We assess the generalization ability of \sys through \textbf{\sys-Transfer}, a cross-dataset transfer setting within identical task categories. We first train \sys on a source dataset to obtain trained agents, then directly use these agents to generate a cleaning pipeline on a target dataset of the same task type, and finally evaluate the cleaned target dataset. Specifically, for forecasting, the agents trained on ETTh1 are transferred to IDF\_OilTemp to generate cleaning pipelines, and the cleaned data is then evaluated, while the reverse transfer is conducted analogously. The same source-to-target transfer protocol is applied to Libras and Handwriting for classification and clustering.

%ETTh1 and IDF\_OilTemp are used as paired datasets: 

Table~\ref{tab:generalization_evaluation} evaluates the cross-dataset generalization of \sys by examining the interplay between cleaning effectiveness ($NMSE$) and downstream utility ($\Delta\text{Perf}$). Across all tasks, the consistently low $NMSE$ (ranging from 0.0012 to 0.0085)  indicates that the transferred agents can generate cleaning pipelines that preserve the main temporal patterns of the target data without introducing severe distortions. This zero-shot fidelity stems from the MDP design, where the agent makes decisions based on dataset-agnostic distributional and structural features rather than raw numerical values. Furthermore, the predominantly positive $\Delta\text{Perf}$ scores across transfer scenarios, peaking at a +0.1589 gain (ETTh1 $\rightarrow$ IDF\_OilTemp) with only a negligible -0.0034 dip (Handwriting $\rightarrow$ Libras), suggest that \sys can transfer quality-aware cleaning policies across datasets within the same task category to actively boost downstream utility on novel datasets in a zero-shot manner. This is because the agents capture task-level and quality-level cleaning patterns, preserving task-relevant discriminative patterns instead of applying blind numerical smoothing, which aligns with the fundamental requirements of downstream tasks across datasets.

This setting is meaningful because training RL-based cleaning agents from scratch for each new dataset can be costly. We therefore examine whether a policy trained on one dataset can be reused on another dataset with the same task type and similar multivariate time-series modality. This does not assume identical physical semantics or numerical distributions across domains; rather, the transferred knowledge lies in quality-aware cleaning strategies learned from data-quality indicators and downstream feedback.

\begin{figure}[!t]
  \centering
  \includegraphics[width=\linewidth, height=0.9\linewidth, keepaspectratio]{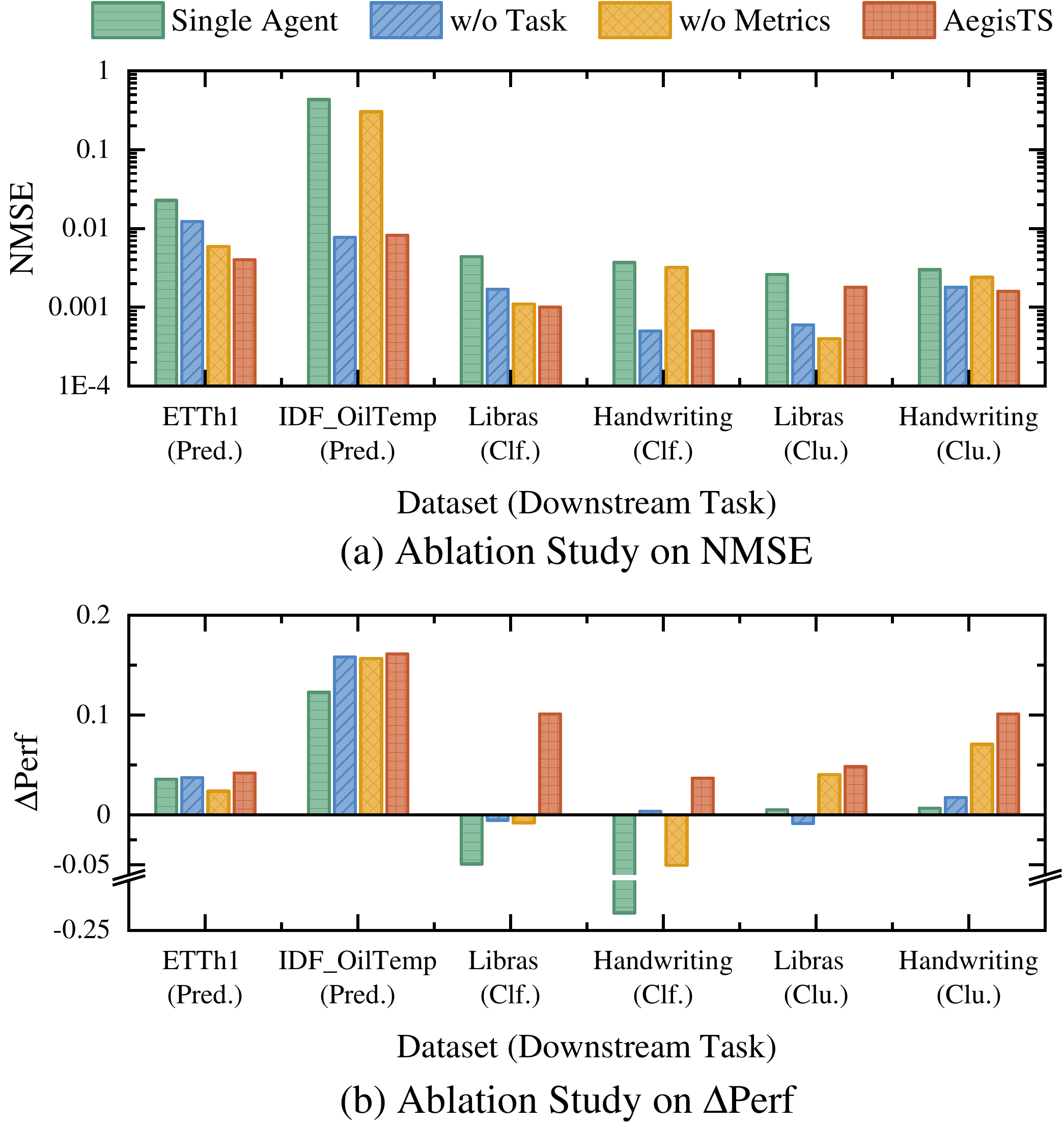}
  \vspace{-3mm}
  \caption{Ablation study.}
  \vspace{-3mm}
  \Description{Ablation.}
  \label{fig:ablation}
\end{figure}

\begin{figure*}[!t]
  \centering
  \includegraphics[width=0.95\textwidth,height=0.8\linewidth,keepaspectratio]{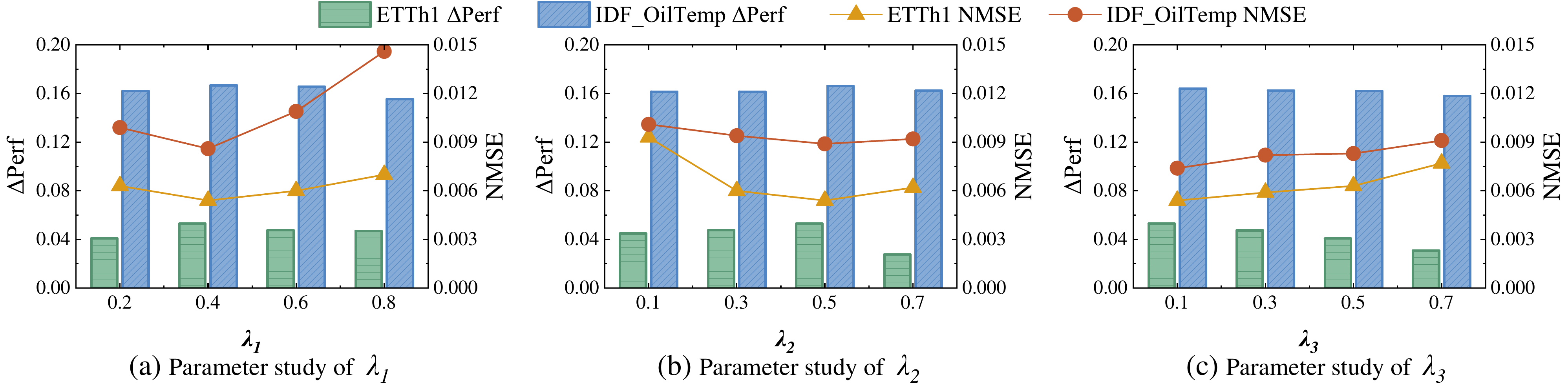}
  \vspace{-4mm}
  \caption{Parameter study of \textbf{$\lambda$} values.}  
  \Description{High-level reward parameters.}
  \vspace{-2mm}
  \label{fig:lambda}
\end{figure*}

\begin{figure}[!t]
  \centering
  \includegraphics[width=\linewidth,height=\linewidth,keepaspectratio]{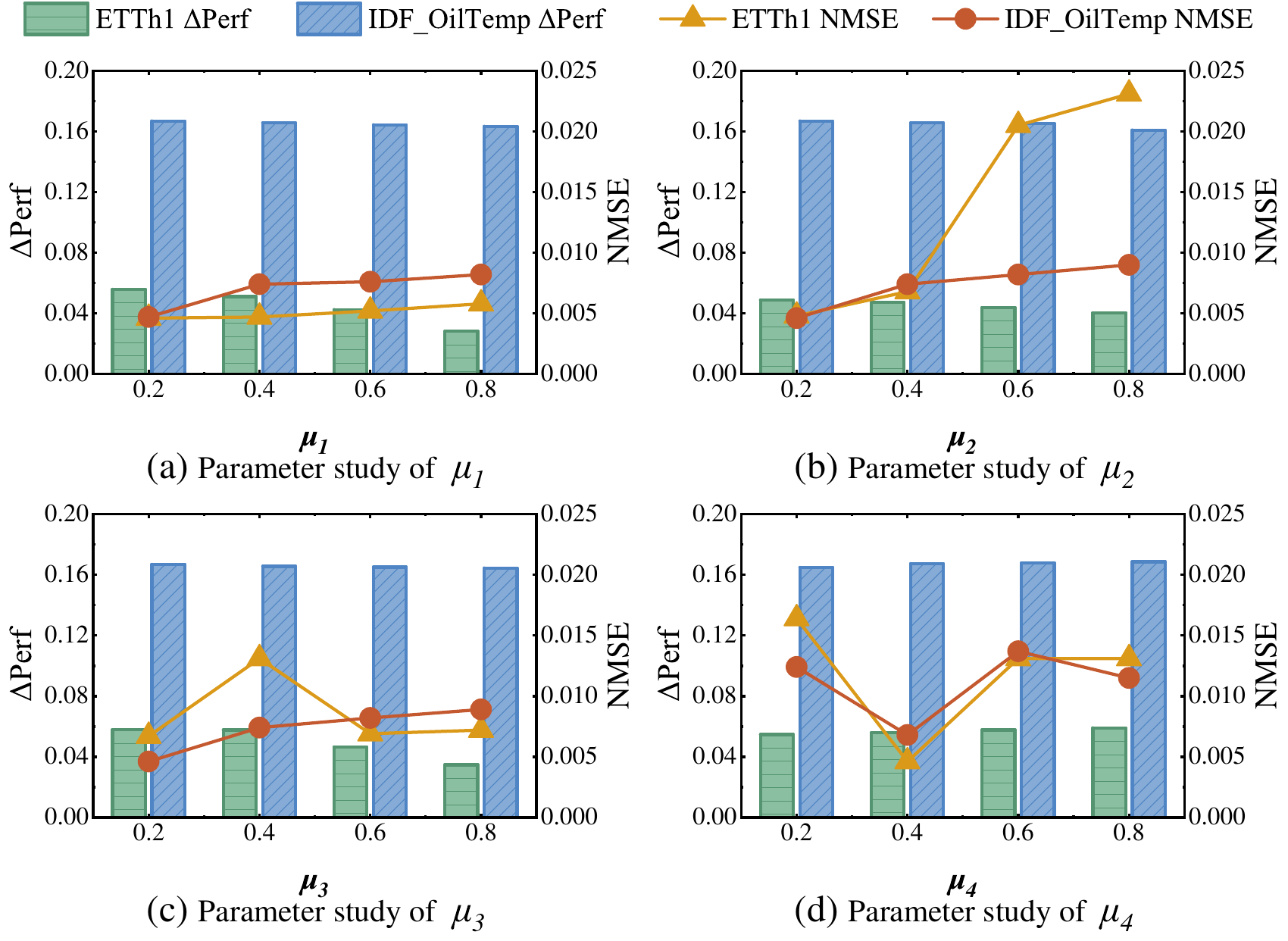}
  \vspace{-6mm}
  \caption{Parameter study of \textbf{$\mu$} values.}
  \vspace{-4mm}
  \Description{Low-level reward parameters.}
  \label{fig:mu}
\end{figure}

% \begin{figure}[!t]
%   \centering
%   \includegraphics[width=\linewidth, keepaspectratio]{figures/ablation.pdf}
%   \caption{Ablation Evaluation.}
%   \Description{Ablation.}
%   \label{fig:ablation}
% \end{figure}

\subsection{Efficiency Evaluation}

The time cost of different methods is shown in~\autoref{fig:time}. 
Brute-Force Search is omitted because it takes more than three days on all datasets.
Sampling and EDITOR are relatively fast on several datasets. However, as shown by the downstream-task results in Table~\ref{tab:comprehensive_evaluation}, their efficiency is generally accompanied by lower downstream utility than \sys. 
Although \sys incurs additional training overhead, it remains more efficient than the flat `Single Agent' variant across all tasks, indicating that hierarchical decision-making effectively narrows the search space. 
Combined with its superior downstream performance reported earlier, these results demonstrate that \sys achieves a favorable balance between computational efficiency and cleaning effectiveness. 
On IDF\_OilTemp, \sys is even faster than Clean4TSDB, EDITOR, and Sampling due to the dataset's low dimensionality, strong inter-attribute correlations, and regular temporal patterns, which favor lightweight modeling and accelerated convergence.

From a practical perspective, the generalization ability of \sys enables further acceleration through \sys-Transfer. 
As shown in~\autoref{fig:time}, \sys-Transfer substantially reduces runtime by reusing agents trained on a source dataset to generate cleaning pipelines for target datasets within the same task category, without additional retraining. 
Together with its competitive downstream performance reported in Table~\ref{tab:generalization_evaluation}, \sys-Transfer provides a practical balance between deployment efficiency and cleaning utility.

\subsection{Ablation Study}
We perform ablation studies on three key components to quantify their contributions to \sys, as illustrated in ~\autoref{fig:ablation}. Specifically, \textbf{Single Agent} replaces the hierarchical architecture with a single flat agent that directly searches the full operator space. \textbf{w/o Task} removes downstream task–guided rewards, retaining only task-agnostic reward terms. \textbf{w/o Metrics} excludes reward signals derived from quality issue rates in the Quality Evaluator (i.e., $\mathbf{r}_{missing}$, $\mathbf{r}_{outlier}$, and $\mathbf{r}_{violation}$ in Section~\ref{sec:quality_rates}). 

The results in ~\autoref{fig:ablation} reveal three key observations. First, the \textbf{Single Agent} variant suffers from severe performance degradation, yielding negative performance gains in classification ($\Delta\text{Perf}<0$) and substantially increased $NMSE$. This indicates that flattening the hierarchical action space exposes the RL agent to excessive combinatorial complexity, whereas the proposed hierarchical design enables more efficient and safer policy exploration. Second, the \textbf{w/o Task} variant fails to consistently achieve the lowest $NMSE$ because it optimizes solely for statistical data quality. This highlights the role of downstream task feedback as a crucial structural prior that preserves discriminative temporal patterns by preventing over-smoothing. Third, the \textbf{w/o Metrics} variant suffers from severe $NMSE$ spikes and limited performance gains, as the absence of explicit quality issue indicators may lead the agent to subtly distort the data within conservative cleaning bounds to overfit downstream model preferences. These results confirm that the Quality Evaluator provides indispensable fine-grained supervision, guiding cleaning policies to restore the underlying data structure rather than generate task-overfitted artifacts.

The ablation results illustrate the complementary roles of the hierarchical agent architecture and the reward formulation that integrates downstream task feedback with explicit quality issue rates. Together, they enable \sys to effectively navigate the cleaning space and produce high-fidelity cleaned data.
%that consistently enhances downstream task utility.

% \vspace{-5mm}

\subsection{Parameter Study}
Finally, we evaluate the sensitivity of reward weights. \autoref{fig:lambda} shows results for the high-level parameters $\lambda$ (Equation~\ref{eq:high_dense}), while \autoref{fig:mu} shows results for the low-level parameters $\mu$ (Equation~\ref{eq:low_dense}). Overall, IDF\_OilTemp is relatively stable across different weights, while ETTh1 is more sensitive in NMSE under several low-level parameters; thus, we use moderate default weights to avoid extreme reward preferences.

%{\textbf{High-Level Parameters ($\lambda$):} }
\softbsubsec{High-Level Parameters ($\lambda$)}
1) \textit{Execution feedback propagation} ($\lambda_1$): $\Delta$Perf first increases and peaks at $0.4$, while $NMSE$ reaches a relatively low value, and then performance declines. This is because a moderate $\lambda_1$ introduces useful downstream feedback, whereas a larger value leads the policy to overfit local rewards and weaken global structural constraints.
2) \textit{Global quality improvement} ($\lambda_2$): Performance is optimal at $\lambda_2 = 0.5$, after which $\Delta$Perf drops. This is because overly strong global constraints induce over-smoothing, which damages discriminative patterns.
3) \textit{Computational cost penalty} ($\lambda_3$): As $\lambda_3$ increases, $NMSE$ rises and $\Delta$Perf declines, and we use a small penalty of $0.1$ as the default setting. This is because a higher cost penalty encourages premature termination, resulting in under-cleaned data.
%As $\lambda_3$ increases, $NMSE$ rises and $\Delta$Perf declines, with the best value at $0.1$. This is because a higher cost penalty encourages premature termination, resulting in under-cleaned data.

\softbsubsec{Low-Level Parameters ($\mu$)}
%{\textbf{Low-Level Parameters ($\mu$):} }
1) \textit{Temporal smoothness} ($\mu_1$): As $\mu_1$ increases, $\Delta$Perf steadily declines and $NMSE$ rises, with the best value at $0.2$. This is because excessive smoothness constraints prevent necessary corrections, leaving severe errors under-repaired.
2) \textit{Modification constraint} ($\mu_2$): Larger $\mu_2$ leads to higher $NMSE$ and lower $\Delta$Perf on ETTh1, while the results are more stable on IDF\_OilTemp. We set $\mu_2=0.2$ as a conservative default, since excessive penalties restrict the repair of noise and anomalies.
%Larger $\mu_2$ leads to higher $NMSE$ and lower $\Delta$Perf, with $0.2$ performing best, due to excessive penalties restricting the repair of noise and anomalies.
3) \textit{Local effectiveness} ($\mu_3$): Increasing $\mu_3$ causes only mild changes in $\Delta$Perf but leads to unstable $NMSE$ on ETTh1. We use $\mu_3=0.2$ to balance local repair effectiveness and global structure preservation.
%Increasing $\mu_3$ reduces $\Delta$Perf and causes unstable $NMSE$, making $0.2$ optimal, as the agent becomes biased toward point-wise fixes that weaken global structure.
4) \textit{Task alignment} ($\mu_4$): $NMSE$ shows a V-shaped trend and reaches its minimum at $\mu_4=0.4$, while $\Delta$Perf remains relatively high. This is because moderate task alignment guides useful repairs, whereas excessive task emphasis may introduce unnatural artifacts and increase reconstruction error.

\section{Related Work}
\label{sec:rel_work}

\subsection{Time Series Data Cleaning Methods}
Existing studies on time series data cleaning primarily focus on methods for detecting and repairing erroneous observations. These approaches can be broadly categorized into three families.
%: statistical, constraint, and anomaly detection-based methods.

% \textbf{Smoothing-based methods} reduce noise by estimating underlying trends and repairing suspicious points accordingly. Moving Average (MA)~\cite{DBLP:books/daglib/0005327} replaces each value with neighboring averages, while Exponentially Weighted Moving Average (EWMA)~\cite{gardner1985exponential} emphasizes recent observations through decaying weights. Autoregressive (AR)~\cite{DBLP:journals/tkde/TakeuchiY06}, ARMA~\cite{DBLP:conf/icassp/AlengrinF78}, and ARIMA~\cite{box1970distribution} model temporal dependencies through linear autoregression, residual correlations, and differencing. These methods are efficient and easy to implement, but may modify correct observations and often assume stable linear patterns, limiting their effectiveness on highly dynamic or nonlinear time series.

\textbf{Statistics-based methods} model time series probabilistically and detect errors as deviations from expected distributions. Maximum Likelihood Estimation (MLE)~\cite{DBLP:journals/tsp/BreslerM86} flags low-likelihood points by estimating parameters that maximize data likelihood, while Bayesian models~\cite{DBLP:conf/icml/GetoorFKT01} incorporate priors and update posterior beliefs for repair. Markov models~\cite{dukhovny1990markov} capture temporal state transitions, and Hidden Markov Models (HMM)~\cite{gupta2012stock} introduce latent states for likelihood-based inference. Similarly, MissNet~\cite{DBLP:conf/kdd/ObataKMS24} uses state-space models to mine latent dynamics and infer missing values. Expectation--Maximization (EM)~\cite{shumway1982approach} estimates hidden parameters iteratively, while Akane~\cite{DBLP:journals/pacmmod/HanXHWWW24} selects the most probable sequence under a learned probabilistic model. These methods capture temporal dependencies, but often require strong assumptions and sufficient historical data, limiting their robustness to nonlinear, highly dynamic, or non-stationary patterns.

\textbf{Constraint-based methods} detect and repair errors by enforcing predefined rules derived from domain knowledge. SCREEN~\cite{DBLP:conf/sigmod/SongZWY15} uses speed constraints to identify and correct temporal anomalies, while SpeedAcc~\cite{DBLP:journals/tods/SongGZWY21} further incorporates velocity and acceleration constraints for consistency checking. 
%Yin et al.~\cite{DBLP:conf/dasfaa/YinYWHL18} apply variance constraints to detect outliers with abnormal variability.
For multivariate time series, MTSClean~\cite{DBLP:journals/pvldb/DingSWWY24} and Clean4MTS~\cite{DBLP:conf/icde/DingLWWS24} combine row and column constraints for comprehensive error detection and repair. MTCSC~\cite{DBLP:journals/pacmmod/ZhangWGYW24} integrates clustering to improve repair accuracy, while Cleanits~\cite{DBLP:journals/pvldb/DingWSLLG19} addresses missing values, outliers, and structural inconsistencies through statistical correlation modeling and sequence constraints. Targeting time series databases, Clean4TSDB~\cite{DBLP:journals/pvldb/DingSWYWW24} introduces context-aware dependency constraints to capture broader structural relationships. Overall, constraint-based methods offer high interpretability and consistency guarantees by exploiting prior knowledge, but their performance depends heavily on constraint design and may suffer from high computational cost and limited scalability on complex or large-scale datasets.

\textbf{Anomaly Detection-based methods} treat cleaning as anomaly detection followed by repair. Distance-based methods~\cite{DBLP:journals/pvldb/DingWSLLG19} flag points far from their neighbors, while clustering-based methods~\cite{DBLP:conf/kdd/EsterKSX96} identify points that do not fit any cluster. Deep learning approaches further capture complex temporal patterns: LSTM models~\cite{DBLP:conf/esann/MalhotraVSA15} detect anomalies via reconstruction errors, GAN-based methods~\cite{DBLP:conf/itsc/SunPLS18} learn data distributions adversarially, TranAD~\cite{DBLP:journals/pvldb/TuliCJ22} models long-range dependencies with transformers, and ImDiffusion~\cite{DBLP:journals/pvldb/ChenZMLDLHRLZ23} uses diffusion-based reconstruction for anomaly detection. EDITOR~\cite{li2026editor} enables context-aware repair through a bidirectional framework combining TCN-based temporal modeling and GCN-based inter-attribute interaction learning. These methods are effective for subtle anomalies and nonlinear dynamics, but their repair quality depends heavily on detection accuracy.
%and sufficient training data.

Existing methods target isolated quality issues and struggle with co-occurring errors, while naive chaining can break temporal and cross-attribute dependencies. We propose \sys, which treats these methods as \textit{cleaning operators} and formulates cleaning as a sequential decision process. Using hierarchical reinforcement learning, \sys dynamically selects and orders operators to construct a dataset-specific \textit{cleaning pipeline} without clean ground truth.

\subsection{Time Series Data Cleaning Systems}
Several systems have been proposed to automate data cleaning using different technical paradigms. \textit{DiffPrep}~\cite{DBLP:journals/pacmmod/LiCC023} models data preparation as a differentiable process and jointly optimizes cleaning operations with downstream model training via gradient-based learning; however, it requires the downstream task to be differentiable, limiting its applicability in many practical scenarios. \textit{DIANA}~\cite{sancricca2025lightweight} supports lightweight data preparation for tabular data by prioritizing influential quality issues, aiming to reduce preparation cost while preserving analysis quality. \textit{HoloClean}~\cite{DBLP:journals/pvldb/RekatsinasCIR17} formulates data repairing as a probabilistic inference problem by integrating integrity constraints, statistical signals, and external knowledge sources, but its effectiveness heavily depends on the availability and correctness of predefined constraints. \textit{AutoDCWorkflow}~\cite{DBLP:conf/emnlp/LiFLT25} leverages large language models to automatically construct data cleaning workflows, yet its reliance on LLM inference incurs substantial token costs and latency, making it less suitable for large-scale or low-latency settings. \textit{ReClean}~\cite{DBLP:conf/icde/AbdelaalYKS24} adopts RL to learn data cleaning strategies through iterative interaction and feedback.

Most of these systems focus on tabular data and ignore temporal dependencies. Instead, we propose a hierarchical RL system for multivariate time series cleaning. \sys formulates cleaning as sequential decision-making over an operator repository, guided by a dual-stage reward combining quality signals and task performance.

\vspace{-2mm}
\section{Conclusion}
\label{sec:conclusion}

% We present \sys, a hierarchical reinforcement learning system for automated cleaning of multivariate time series with multiple quality issues. It separates decision-making into a high-level policy for prioritizing issues and a low-level policy for selecting cleaning operations, enabling efficient exploration of large cleaning pipelines while preserving temporal and cross-attribute structure.
% To overcome the lack of ground truth, we design a dual-stage reward combining step-wise data quality improvements with end-task performance feedback. Experiments on real-world datasets show that \sys outperforms existing methods in data quality, downstream accuracy, and efficiency.
% Future work will extend \sys to streaming data and incorporate time series foundation models to improve generalization to unseen domains.

We present \sys, a hierarchical reinforcement learning system for automated cleaning of multivariate time series with multiple quality issues. It decomposes cleaning strategy construction into high-level issue prioritization and low-level operator selection, enabling efficient exploration of large cleaning pipelines. 
To reduce the reliance on clean ground truth, \sys mines constraints directly from dirty data and combines quality-related signals with downstream feedback in a dual-stage reward design. 
Experiments on datasets show that \sys outperforms existing methods in data quality, downstream accuracy, and efficiency. 
Although the hierarchical design reduces the search space, \sys still incurs additional training and evaluation overhead compared with lightweight cleaning baselines.
In the future, we will focus on reducing this overhead and extending \sys to streaming scenarios. %, where online state updates, incremental constraint maintenance, low-latency operator selection, and concept-drift-aware policy updating are required under limited historical context.

\balance
\bibliographystyle{ACM-Reference-Format}
\bibliography{references}

\appendix

\end{document}